\documentclass[useAMS,usenatbib,referee]{mn2e}

\usepackage{graphicx,color}

\newcommand{\etal }{{et al.} }
\newcommand{\msun}{\thinspace M_\odot}

\newcommand{\rsun}{\thinspace R_\odot} 
\newcommand{\zsun}{\thinspace Z_\odot} 
\newcommand{\vect}[1]{\mbox{\boldmath$#1$}}
\def\lesssim{\mathrel{\hbox{\rlap{\hbox{\lower4pt\hbox{$\sim$}}}\hbox{$<$}}}}
\def\gtrsim{\mathrel{\hbox{\rlap{\hbox{\lower4pt\hbox{$\sim$}}}\hbox{$>$}}}}
\newcommand{\cm}{\,{\rm cm}^{-3} }

\newcommand{\tps}{t_{\rm ps}}

\newcommand{\dfrac}[2]{{\displaystyle \frac{#1}{#2}} }

\title[Different Metallicities]{Accretion Phase of Star Formation in Clouds with Different Metallicities}
\author[M.N.~Machida \& T.~Nakamura ]
  { Masahiro N.~Machida$^{1}$\thanks{E-mail: machida.masahiro.018@m.kyushu-u.ac.jp (MNM)} and Teppei Nakamura$^{1}$  \\
$^{1}$Department of Earth and Planetary Sciences, Faculty of Sciences, Kyushu University, 6-10-1 Hakozaki, Higashi-ku, Fukuoka, Japan\\
}

\begin{document}
\maketitle
\begin{abstract}
The main accretion phase of star formation is investigated in clouds with different metallicities in the range of $0\le Z \le \zsun$, resolving the protostellar radius. 
Starting from a near-equilibrium prestellar cloud, we calculate the cloud evolution up to $\sim100$\,yr after the first protostar formation.
The star formation process considerably differs between clouds with lower ($Z\le 10^{-4}\zsun$) and higher ($Z>10^{-4}\zsun$) metallicities.
Fragmentation frequently occurs and  many protostars appear without forming a stable circumstellar disc in lower-metallicity clouds. 
In these clouds, although protostars mutually interact and some are ejected from the cloud centre, many remain as a  small stellar cluster. 
In contrast, higher-metallicity clouds produce a single protostar surrounded by a nearly stable rotation-supported disc.   
In these clouds, although fragmentation occasionally occurs in the disc, the fragments migrate inwards and finally fall onto the central protostar. 
The difference in cloud evolution is due to different thermal evolutions and mass accretion rates. 
The thermal evolution of the cloud determines the emergence and lifetime of the first core. 
The first core develops prior to the protostar formation in higher-metallicity clouds, whereas no (obvious) first core appears in lower-metallicity clouds. 
The first core evolves into a circumstellar disc with a spiral pattern, which effectively transfers the angular momentum outwards and suppresses frequent fragmentation.
In lower-metallicity clouds, the higher mass accretion rate increases the disc surface density within a very short time, rendering the disc unstable to self-gravity and inducing vigorous fragmentation.

\end{abstract}

\section{Introduction}
\label{sec:intro}
Stars control the dynamical and chemical evolution of the universe. 
The typical stellar mass, stellar feedback and final fate of stars are closely related to the evolution of the universe, and the latter two factors are determined by the stellar mass at the moment of stellar birth.
Thus, it is very important to understand the star formation processes throughout the history of the universe.
Recent studies have shown that the star formation process in primordial gas clouds considerably differs from that in present-day clouds and in clouds formed in the middle-aged universe \citep{bromm01,smith07,jappsen07,machida08s,clark08,jappsen09,smith09,machida09a, dopcke11,dopcke13,safranek14}. 
A major difference between primordial and present-day clouds is the abundance of metals and dust.
Primordial gas clouds contain neither metals nor dust, while present-day clouds contain both. 
The abundance of metals and dust influences the thermal evolution of collapsing star-forming clouds \citep{omukai00,omukai05,omukai10}.

A collapsing gas cloud with a primordial composition cannot cool to below $\sim 200$\,K, because, except for molecular hydrogen, there is no effective coolant at temperatures as low as  $T\lesssim 200$\,K.
On the other hand, a present-day gas cloud can cool to $\sim 10$\,K because cooling by the metal and dust is effective even at $T\lesssim 200$\,K.  
Numerical simulations have shown that star-forming clouds (or mini-halos) are formed at a number density of $n_{\rm H} \sim10^{4}-10^{5}\cm$ in the early universe \citep[e.g.][]{abel02, bromm02,yoshida06}, while observations of nearby star-forming regions have shown that molecular cloud cores have a number density of $n_{\rm H_2}\sim10^{4}-10^{5}\cm$ \citep[e.g.][]{bergin07}. 
Thus, star-forming clouds have almost the same density of $n\sim10^4-10^5\cm$ in both the early and present-day universes.
However, these clouds have different temperatures with different abundances of metals and dust \citep{omukai05}.
With a cloud density of  $n\sim10^4\cm$, primordial and present-day clouds have temperatures of $\sim200$\,K and $\sim10$\,K, respectively. 
If an equilibrium state is assumed to be realized in a star-forming cloud, the cloud mass depends on the gas temperature (or metallicity) under the assumption of spherical symmetry. 
Reflecting the difference in the gas temperature in clouds, star-forming clouds at each age should have different masses.
For example, primordial star-forming clouds have a mass of $\sim10^3\msun$ \citep{bromm02}, while present-day clouds (molecular cloud cores) are  $\sim 1\msun$ \citep{andre10}. 

After protostar formation in a collapsing  cloud, the mass accretion rate onto the protostar also depends on the gas temperature of the star-forming cloud.
Theoretically, the mass accretion rate is expected to be proportional to $\sim c_s^3/G$ \citep{shu77,larson03}, where $c_s$ is the sound speed of the prestellar cloud. 
Thus, the mass accretion rate should be approximately 100 ($\sim (200{\rm K}/10{\rm K})^{1.5}$) times larger in a primordial cloud than in a present-day cloud. 
Therefore,  primordial stars are considered to be more massive than  present-day stars (see also \citealt{omukai01,omukai03}).
However,  this rough estimate ignores fragmentation in the collapsing cloud. 
A single massive star forms without fragmentation, whereas vigorous fragmentation leads to a multiple stellar system comprising less massive stars.

We cannot directly observe the formation sites of stars even in present-day clouds because they are embedded in  dense gas envelopes. 
Thus, the star formation process must be clarified by a theoretical approach and numerical simulations. 
Several numerical simulations have investigated the evolution of primordial and present-day star-forming clouds. 
However, simulating a collapsing cloud over sufficiently long durations to resolve the protostar is very difficult, because the numerical timestep shortens as the gas density increases.
Since the protostar has a very high density, cloud evolution around a protostar requires an extremely short timestep precluding long-term integration. 
The sink method is used to overcome this problem, in which the high-density gas is assumed to become a protostar and is replaced by sink particles or sink cells \citep{bate95,krumholz04,hubber13}.
At the expense of spatial resolution inside and near the sink accretion radius, some studies have realized the long-term simulation of star-forming clouds with different metallicities \citep{bromm01,clark08,dopcke11,dopcke13,safranek14}.
Although these studies have provided insight into large-scale fragmentation, star formation efficiency and stellar mass distribution, the evolution of the circumstellar region is masked by the sink.
In the protostar proximity, we can expect small-scale circumstellar disc formation and fragmentation, which may affect the large-scale cloud evolution.

Some studies have investigated protostar formation in present-day star-forming clouds without employing a sink \citep{bate98,machida07,bate11,machida11a,tomida13,bate14}. 
These studies could model cloud evolution for only $\sim$1-100\,yr after the protostar formation, and demonstrated that fragmentation is rare in high-density gas regions.
In addition, these studies indicate that a rotation-supported disc forms {\it before} the protostar formation, where the first core \citep{larson69,masunaga00} formed before the protostar formation evolves into a circumstellar disc.
Thus, at the protostar formation epoch, the protostar is already surrounded by a large-scale rotation-supported disc with spiral structures, which effectively transfer angular momentum and stabilize the disc against gravitational instability and subsequent fragmentation.

On the other hand, the evolution of primordial star-forming clouds on protostar-resolving scales has been scarcely investigated in models without a sink  \citep{greif12,machida13a}.  
\citet{greif12} showed that fragmentation is common around initially formed primordial protostars. 
\citet{machida13a} reported  that no stable circumstellar disc is established around a primordial protostar because vigorous fragmentation and the orbital motion of the fragments significantly disturb the circumstellar environment.
They also showed that when a sink is introduced into a primordial cloud, the fragmentation scale is (artificially) determined by the sink accretion radius. 
Furthermore, in primordial star-forming clouds, a rotation-supported disc cannot form before the protostar formation because the first core does not appear in the primordial collapsing gas \citep{omukai00, omukai05, yoshida06,greif12,greif13,machida13a}.

Previous studies without a sink reveal significant differences between the primordial and present-day star formation processes. 
Primordial clouds frequently fragment without forming a stable circumstellar disc, whereas the stable rotation-supported disc prevents frequent fragmentation in present-day clouds. 
Thus, it is considered that the star formation process depends upon the cloud metallicity. 
The  large-scale cloud evolution after the protostar formation at different metallicities has been widely investigated with a sink, but has not been investigated without a sink.
Although a sink may be required in long-term cloud evolution calculations, we should first study the cloud evolution by resolving the protostar to investigate the effect of the small-scale structure around the protostar on the large-scale structure.
In this study, we adopt a protostellar model similar to that of \citet{machida13a} and investigate the star formation process from the prestellar cloud stage until approximately 100\,yr after the protostar formation, varying the cloud metallicities within the range of $0\le Z \le \zsun$. 

The rest of the paper is structured as follows.
Our model framework and numerical methods are described in \S 2 and the protostellar model is explained in \S 3. 
The numerical results are presented in \S 4, 
while \S 5 is devoted to the fragmentation condition and effects of the initial condition.  The results are summarized  in \S 6.

\section{Model and Numerical Method}
\label{sec:model}
We calculate the evolution of eight clouds with different metallicities in the range of $0 \le Z \le \zsun$.
The initial clouds have the critical Bonner--Ebert (BE) density profile, which is characterized by the central density and isothermal temperature \citep{ebert55,bonnor56}. 
To construct the  BE profile, we set the central density as $n=10^4\cm$ in all models and adopt different isothermal temperatures that are estimated by a one-zone calculation (see below). 
To promote contraction, we enhance the cloud density by a factor of 1.8 \citep{matsu03}. 
For each initial cloud, the mass and radius are different, while the initial cloud stability (the ratio of thermal to gravitational energy), which affects the fragmentation condition, is the same (for details, see \S\ref{sec:init}). 
The cloud metallicity $Z$, initial (central) temperature $T_{\rm cl}$, cloud mass $M_{\rm cl}$ and radius $R_{\rm cl}$ in each model are listed in Table~\ref{table:1}. 

For all models, a rigid rotation of $\Omega_0=3\times10^{-15}$\,s$^{-1}$ is adopted.
Each initial cloud has the same ratio of rotational to gravitational energy of $\beta_0=10^{-3}$. 
Although the rotational energy adopted in this study lies within the observational constraints of nearby star-forming regions (e.g. $10^{-4}<\beta_0<0.07$; \citealt{caselli02}), it may be somewhat smaller than that estimated by recent cosmological simulations \citep[e.g.][]{hirano14}. 
We adopt the same value of $\beta_0$ for all models to limit calculation models, while the initial rotation rate may influence the fragmentation process \citep[e.g.][]{machida08}. 
In addition, the initial cloud mass may also be important for investigating the star formation process. 
The effects of the initial rotational energy and cloud mass  are discussed in \S\ref{sec:init}.

The cloud evolution is computed by the nested grid method \citep[for details, see][]{machida04,machida05a,machida05b}, which constructs nested rectangular grids with the same number of cells (i, j, k) = (256, 256, 32) or (128, 128, 16). 
We impose a mirror symmetry with respect to the $z=0$ plane. 
Initially, we prepare the fifth grid level and immerse the initial cloud in the first grid level, whose box size is twice the initial cloud radius $2R_{\rm cl}$. 
A fixed boundary condition is imposed on the surface in the first grid level. 
As the calculation proceeds, finer grids are dynamically generated, which resolve the Jeans length in at least 16 cells \citep{truelove97}. 
Although the grid sizes and cell widths differ between the models even at the same grid level $l$, we can appropriately calculate the cloud evolution and fragmentation process because the Truelove condition is always fulfilled (except for the escaped fragments, see \S\ref{sec:results}).

Our numerical method solves the following hydrodynamics equation with  self-gravity:
\begin{eqnarray} 
& \dfrac{\partial \rho}{\partial t}  + \nabla \cdot (\rho \vect{v}) = 0, & \\
& \rho \dfrac{\partial \vect{v}}{\partial t} 
    + \rho(\vect{v} \cdot \nabla)\vect{v} =
    - \nabla P -      \rho \nabla \phi, & 
\label{eq:eom} \\ 
& \nabla^2 \phi = 4 \pi G \rho, &
\end{eqnarray}
where $\rho$, $\vect{v}$, $P$ and $\phi$ denote the density, velocity, pressure and gravitational potential, respectively. 
For gas pressure, we adopt a barotropic relation from the one-zone calculation given by \citet{omukai05}. 
Their one-zone model calculates the thermal evolution of collapsing gas clouds with different metallicities ($0\le Z \le \zsun$).
In Figure~\ref{fig:1}, the thermal evolution of a collapsing gas cloud at different metallicities is plotted against the number density. 
An almost identical barotropic relation was used in our previous studies \citep{machida08s,machida09a}.  
In the present study, we adopt a slightly different thermal evolution. 
Specifically, we impose a lower temperature limit of 10\,K to model a more realistic scenario. 
In previous studies, the gas in clouds with $Z=\zsun$ cools to $\sim3$\,K in the range of $10^4\cm \lesssim n \lesssim 10^{10}\cm$ \citep{machida08s,machida09a}.
Note also that since our present study focuses on high-density gas and the early star formation process, this difference does not significantly affect the results. 

For each metallicity,  two evolutionary tracks are plotted in Figure~\ref{fig:1}.
One track is given by the one-zone calculation, which yields a smoothly connected evolution from low- to high-density gas. 
In the other track, the protostellar model is added to the original one-zone result. 
The protostellar model visibly affects the results only at high density ($n \gtrsim 3 \times 10^{17}\cm$). 
The polytropic index $\gamma$ of the original one-zone calculation is $\sim1.1$ in the range of $10^{16}\cm \lesssim n \lesssim 10^{20}\cm$, whereas that of the protostellar model is $\sim4$ in the high-density gas region. 
The protostellar model is described in the next section.

\section{Protostellar Model}
\label{sec:psmodel}
In many previous studies, the sink method is introduced to investigate long-term cloud evolution, in which the high-density gas region is masked by sink cells or sink particles and is not spatially resolved. 
However,  as described in \S\ref{sec:intro}, a sink tends to mislead the result because, for example, the fragmentation scale is artificially determined by the sink accretion radius \citep{machida13a}.  
To avoid artificial effects caused by a sink and to investigate the early star formation process in a high-density gas region, we here use a simple protostellar model, the same as that adopted in \citet{machida13a}, where the thermal evolution or the polytropic index is adjusted in the high-density gas region and the protostellar radius is related to the protostellar mass.
 
To connect the protostellar mass and radius, we refer to the protostellar mass-radius relation demonstrated  by \citet{omukai10}, 
who used their one-zone calculation to estimate the mass accretion rate onto the protostar in clouds with different metallicities. 
In our study, we first check the long-term mass accretion rate by setting up non-rotational equivalents of the initial clouds listed in Table~\ref{table:1} and by calculating them {\em with a sink}. 
We set the sink accretion radius to $r_{\rm acc}=1$\,AU and the threshold density to $n_{\rm thr}=10^{16}\cm$.
Then, we calculate the cloud evolution for $\sim10^3-10^4$\,yr after the protostar formation, in which we define the protostar formation epoch at which the maximum cloud density reaches the threshold density \citep[for details of the sink method, see][]{machida13a}.

The mass accretion rate for different metallicities is plotted against the protostellar mass in Figure~\ref{fig:2}.
In clouds with higher metallicity ($Z>10^{-4}\zsun$), the mass accretion rate is small, and hence, we can calculate the cloud evolution up to a protostellar mass of  $\sim0.1\msun$.  
Although we only calculate the very early phase of star formation, the mass accretion rates in Figure~\ref{fig:2} approximate those reported in \citet{omukai10} (see also \citealt{tanaka14}). 
Since the accretion history mainly governs the protostellar evolution, we expect that, in our calculation, the protostar evolves similarly to the evolutionary process described in  \citet{omukai10}. 
Thus, we can relate the protostellar radius to the protostellar mass by referring  to Figure~17 in \citet{omukai10}.

Next, to construct the protostellar model, we calculate the evolution of a non-rotating cloud {\em without a sink},  as described in  \citet{machida13a}, parameterizing the polytropic index $\gamma_{\rm ps}$ and the protostellar density $n_{\rm ps}$ at which the thermal evolution differs  from the one-zone result.  
For each model listed in Table~\ref{fig:1}, we calculate the protostellar evolution of 5-20 clouds, varying $n_{\rm ps}$ and $\gamma_{\rm ps} $, and determine the most plausible parameters of the protostellar model, with which the mass-radius relation given by \citet{omukai10} are best reproduced. 
Figure~\ref{fig:1} plots the thermal evolution of the protostellar model (red lines) at each metallicity. 
As shown in the figure, the protostellar densities are in the range of $3\times 10^{17} \cm \lesssim n_{\rm ps} \lesssim 3\times 10^{21} \cm $. The polytropic index of $ \gamma_{\rm ps} = 4 $ is adopted for all models. 
The protostellar density in each model is listed in Table~\ref{table:1}.
With a harder equation of state, long-term calculation becomes possible because the density increase almost stops inside the protostar (or inside the shock surface), alleviating the Truelove condition. 
Although we cannot resolve the inner structure of the protostar, we can resolve the circumstellar region just outside the protostar. 
Note that we do not resolve the high-density gas region ($n \gtrsim n_{\rm ps}$) with a stiff EOS, especially for lower-metallicity models ($Z\le10^{-4}\zsun$). 
Although the mass accretion rate is mainly determined by the large scale 
structure without a persistent first core formation in such models \citep{machida09a}, a stiff EOS may somewhat influence the mass accretion rate and fragmentation process.

Figure~\ref{fig:3} plots the protostellar radius against the protostellar mass for different metallicities, in which the cloud evolution is calculated with our protostellar model (Fig.~\ref{fig:1})  and the protostar is defined as the object surrounded by the shock surface in a high-density gas region of $n>3\times10^{17}\cm$. 
In the figure, the protostellar radii for $Z=10^{-5}$, $10^{-4}$, $10^{-3}$, $10^{-1}$ and $1 \zsun$, taken from \citet{omukai10}, are also plotted by thin lines.
The figure shows that our protostellar model reproduces well the mass-radius relation in \citet{omukai10}.
The figure indicates that protostars formed in clouds with $ Z=0$--$10^{-5} \zsun $ have a radius of $ 30 \rsun \lesssim  R_{\rm ps} \lesssim 200 \rsun$ when the protostar has a mass of $M_{\rm ps} <10\msun$. 
On the other hand, the protostellar radii are in the range of $<30\rsun$ for models with $Z\ge10^{-3}\zsun$.
The protostellar mass-radius relations in Figure~\ref{fig:3} are in good agreement with those in \citet{omukai10}.
The protostellar radius increases as $R_{\rm ps} \propto M_{\rm ps}^{1/3}$ during the adiabatic accretion phase.
This phase lasts until the protostar enters the Kelvin-Helmholtz contraction phase \citep{hosokawa09,omukai10,smith11,smith12}.  
As indicated in Figure~\ref{fig:3}, the protostellar radius in each model is roughly proportional to $R_{\rm ps} \propto M_{\rm ps}^{1/3}$. 
Thus, the protostellar model adopted in this study roughly reproduces the protostellar evolution (or the mass-radius relation) reported in previous studies.
{
With this protostellar model (or EOS), we calculate the evolution of clouds with different metallicities. 
To check the protostellar evolution, the detailed evolution of non-rotating clouds is described in Appendix (\S\ref{sec:test}).
}

\section{Results}
\label{sec:results} 
To investigate the effects of metallicity on the star formation process, we prepared prestellar clouds, varying the metallicity in the range of $0\le Z \le \zsun$, and calculated them for $\sim100$\,yr after the first protostar formation (or $\sim10^6$\,yr after the initial cloud begins to collapse).
During the calculations, in any model, fragmentation occurred in the region of $r<20$\,AU and  we could cover the fragmentation region by the finest grid.  
However, some fragments escaped from the central region by mutual gravitational interaction between fragments, and  fragments that had moved far off-centre could not be resolved to sufficiently high spatial resolutions (see \S\ref{sec:model}). 
Thus, we defined the fragments separated by more than $20$\,AU  as `escaped fragments' and ignored their evolution  in subsequent analysis.
Note that although the escaped fragments were calculated with relatively low spatial resolution, they did not significantly affect the cloud dynamics because their mass ($\lesssim 0.1\msun$) was negligible relative to the total cloud mass ($7-2000\msun$).

In the higher-metallicity models, we have to resolve a smaller structure or a higher density region because a protostar formed in a high-metallicity cloud has a smaller size and a higher protostellar density $n_{\rm ps}$, as described in \S\ref{sec:psmodel}. 
To spatially resolve such small protostars, we require a finer grid with smaller cell width around the cloud centre. 
Thus, long-term cloud evolution is limited in these models because a finer resolution requires a shorter time step and much longer CPU  time.
However, for the high-metallicity models of $Z\ge10^{-2}\zsun$, we confirmed that fragmentation is rare and that high-density gas and the protostar(s) occupy the limited region around the cloud centre, as described in the following subsections. 
In such models, a large number of cells are not required in each nested grid to maintain the Truelove condition. 
In the $Z=10^{-2}$, $10^{-1}$ and $1\zsun$ models, we calculated the cloud evolution with cell numbers ($i$, $j$, $k$) = (256, 256, 16) for $\sim30$\,yr after the first protostar formation, and confirmed that fragmentation occurred only in the region near the first formed protostar (or in the circumstellar disc around the protostar).
To reduce the CPU time and  enable long-term calculation,  we then calculated the cloud evolution with cell numbers ($i$, $j$, $k$) = (128, 128, 8),  which is sufficient to maintain the Truelove condition. 
In this setup, we calculated the cloud evolution for over $100$\,yr after the first protostar formation in the $Z=10^{-2}$ and $10^{-1}\zsun$ models, in addition to the $Z=0$, $10^{-6}$, $10^{-5}$, $10^{-4}$ and $10^{-3} \zsun$ models.
The computational time consumed by each model was between four and six months of wall clock time. 
In the $Z=1\zsun$ model, we only calculated the cloud evolution for $55$\,yr after the first protostar formation because a considerably high spatial resolution and considerably  longer calculation time were required.

In this section, we first describe the evolution of each cloud in the main accretion phase during which the protostar significantly increases its mass by mass accretion from the infalling envelope. 
Then, we compare the circumstellar structure, number of fragments and protostellar mass between the different models. 
Although we calculated the collapsing cloud from a prestellar stage with a central density of $10^4\cm$, our main focus is  the cloud evolution after the first protostar formation (i.e. the main accretion phase). 
The evolution of the gas-collapsing phase prior to the  protostar formation has already been investigated in our previous studies \citep{machida06jet,machida08,machida08s,machida08c,machida09a,machida09b}.
In the following subsections, we calculated the cloud evolution {\rm em with} our protostellar model {\em without} a sink.

\subsection{$Z=0$ and $10^{-6}\zsun$ Models}
In this subsection, we briefly describe the cloud evolution in the $Z=0$ and $10^{-6}\zsun$ models. 
The cloud evolution in the $Z=10^{-6}\zsun$ model is almost identical to that in the primordial cloud, which is detailed in \citet{machida13a}.
Figure~\ref{fig:6} shows snapshots of the $Z=0$ (left) and $10^{-6}\zsun$ (right) models at $\tps \sim100$\,yr, where $\tps$ is the elapsed time after the first protostar formation and $\tps=0$ is defined as the epoch at which the maximum density reaches $n_{\rm max}=n_{\rm ps}$ (see Table~\ref{table:1}). 
In the figure, the density (upper panels) and temperature (lower panels) distributions are plotted, in which the temperature distribution is constructed using the density distribution and barotropic equation of state.
Resolving the protostellar scale, \citet{greif12} and \citet{machida13a} showed that collapsing primordial clouds frequently fragment and that the resulting fragments (or protostars) tend to cluster around the centre of the collapsing primordial cloud. 
We refer to a fragment as a protostar when its maximum density in the fragment exceeds the protostellar density, i.e. when  $n_{\rm max}>n_{\rm ps}$, and as a clump when $n_{\rm max}<n_{\rm ps}$.
The upper panels of Figure~\ref{fig:6} show results that are very similar to those of previous studies \citep{clark08, clark11a, clark11b, smith11,greif11,greif12,machida13a}. 
Six fragments are clearly observed in the left panel of  Figure~\ref{fig:6}; the most massive protostar has  a mass of $\sim2\msun$ with a radius of $\sim160\rsun$.
In the $Z=10^{-6}\zsun$ model, fragmentation frequently occurs after the first protostar formation and seven fragments appear, as seen in the Figure~\ref{fig:6} right panel, in which the most massive protostar has a mass of $1.8\msun$ with a radius of $\sim140\rsun$.

In both upper panels, the most massive protostar is surrounded by a  filamentary structure. 
In addition, no stable rotation-supported disc appears after the first protostar formation in either model.
Although a circumstellar disc transiently appears around massive protostars, it immediately fragments into less massive protostars. 
Fragmentation continues in the central region of the cloud and over $\gtrsim 10$ to 20 fragments are observed at most by the end of the calculation.
The cloud evolution during the main accretion phase is essentially the same in the  $Z=0$ and $10^{-6}\zsun$ models because both models have almost identical thermal histories (Fig.~\ref{fig:1}), mass accretion rates (Fig.~\ref{fig:2}) and protostellar mass-radius relations (Fig.~\ref{fig:3}).

The temperature distribution basically traces the density distribution.
However, we can see an interesting feature around some fragments in the left panel ($Z=10^{-6}\zsun$ model): a low-temperature region appears around or inside some fragments.
The low-temperature region corresponds to a temporary temperature drop at $n\sim10^{16}\cm$ (see $Z=0$ evolution track of Figure~\ref{fig:1}). 
This indicates that fragmentation tends to occur in a relatively low-temperature region.

\subsection{$Z=10^{-5}\zsun$ Model}
Figure~\ref{fig:7} shows the time sequence of the density distribution on the equatorial plane around the centre of the collapsing cloud in the $Z=10^{-5}\zsun$ model.
The cloud evolution in this model is qualitatively similar to those in the $Z=0$ and $10^{-6}\zsun$ models; however a slight quantitative difference is evident. 
As shown in Figure~\ref{fig:7}, fragmentation occurs just before the maximum density reaches $n_{\rm ps}$ (Fig.~\ref{fig:7}{\it a}), and two fragments appear (Fig.~\ref{fig:7}{\it b}).
Further fragmentation then occurs around the first formed fragments that have the maximum mass and size (Fig.~\ref{fig:7}{\it c} and {\it d}) among fragments.
For $\tps \gtrsim 50$\,yr, a filamentary structure on a scale of $10-30$\,AU develops and fragments over time. 
As seen in Figures~\ref{fig:7}{\it e}-{\it g}, more than 10 fragments appear within the region of $r<30$\,AU, where $r$ is the radius from the cloud centre.

Less massive protostars tend to fall onto more massive protostars.
Thus, in addition to gas accretion, the massive protostar acquires its mass by mergers with fragments. 
Moreover, less massive protostars are  ejected from the cloud centre by mutual gravitational interaction between protostars.
At the end of the calculation, 11 protostars had been ejected from the cloud centre in the $Z=10^{-5}\zsun$ model. 
Throughout the remainder of the evolution, these protostars moved nearly radially outwards in the region of $r>20$\,AU.
The number of fragments increased up to $\tps \sim 90$\,yr (Fig.~\ref{fig:7}{\it f}-{\it h}) and temporarily decreased thereafter (Fig.~\ref{fig:7}{\it i}).
Fragmentation is ineffective around massive protostars that can stabilize their circumstellar gas against gravitational instability \citep{toomre64}. 
On the other hand, the fragmentation region extends over time because gas with larger angular momentum falls later. 
Since we calculated the cloud evolution for only $\sim 100$\,yr after the first protostar formation, we cannot determine whether vigorous fragmentation continues or whether the system stabilizes.
However, we can conclude that fragmentation is a frequent event and that many protostars appear in the very early phase of star formation in the $Z\le10^{-5}\zsun$ models.  

Figure~\ref{fig:7b} left panel shows the temperature distribution on the equatorial plane at $t_{\rm ps}=60.6$\,yr and indicates that the central region of the collapsing cloud ($r\lesssim 20-40$\,AU) has a lower temperature than the outer envelope ($r\gtrsim 20-40$\,AU). 
The inner low-temperature region corresponds to a temperature decrease in the range of $10^{12}\cm \lesssim n \lesssim 10^{16}\cm$ of Figure~\ref{fig:1}, which is caused by dust cooling \citep{omukai05}. 
A relatively low temperature can cause frequent fragmentation \citep{tanaka14}.

\subsection{$Z=10^{-4}\zsun$ Model}
In the $Z=10^{-4}\zsun$ model, fragmentation continues around the cloud centre until the end of the calculation, but the number of fragments is smaller than that in the $Z\le10^{-5}\zsun$ models. 
Figure~\ref{fig:8} shows the time sequence of the $Z=10^{-4}\zsun$ model. 
The first fragments appear just before the protostar formation (Fig.~\ref{fig:8}{\it a}).
These fragments merge into a single protostar enclosed by a disc or spiral structure (Fig.~\ref{fig:8}{\it b}). 
An additional two fragments appear in the disc, as seen in Figure~\ref{fig:8}{\it c}.
A filamentary structure then develops, with further fragmentation within or near the filament (Figs.~\ref{fig:8}{\it d}-{\it g}).
Some of smaller protostars fall onto more massive protostars, and some are ejected from the cloud centre. 
Only two protostars remain in the region of $r<5$\,AU at $\tps \simeq 70$\,yr (Fig.~\ref{fig:8}{\it g}).
Subsequently, in the evolutionary period of $\tps\sim85-100$\,yr,  only one to three protostars and a few clumps (maximum density of $n_{\rm max} < n_{\rm ps}$) remain.
However, for $\tps \gtrsim100$\,yr, fragmentation is again activated and over five protostars appear in the region of $r<20$\,AU (Fig.~\ref{fig:8}{\it i}).
 Figure~\ref{fig:7b} right panel shows the temperature distribution at $t_{\rm ps}=63.2$\,yr and indicates that  the fragmentation region is embedded in the lower-temperature region caused by dust cooling, as seen in Figure~\ref{fig:7b} left panel. 
Comparing Figures~\ref{fig:6} and \ref{fig:7} with Figure~\ref{fig:8}, the number of fragments is substantially smaller  in the $Z=10^{-4}\zsun$ model than in the $Z<10^{-4}\zsun$ models.


\subsection{$Z=10^{-3}\zsun$ Model}
Unlike the $Z\le10^{-4}\zsun$ models, fragmentation rarely occurs in $Z > 10^{-4}\zsun$ models. 
In the $Z=10^{-3}\zsun$ model, fragmentation occurs in the outer disc region and any fragment falls onto  the central protostar before it can evolve into separate protostars (that is, before its maximum density reaches the protostellar density $n_{\rm ps}$). 
In this model, a single protostar (and a few clumps) remains at the end of the calculation.

Figure~\ref{fig:9} shows the time sequence of the $Z=10^{-3}\zsun$ model. 
As observed in the $Z=10^{-4}\zsun$ model (Fig.~\ref{fig:8}),  the first fragmentation in this model occurs just before the maximum cloud density reaches the protostellar density $n_{\rm ps}$, and two fragments appear (Fig.~\ref{fig:9}{\it a}).
Both fragments evolve into protostars that orbit each other (Figs.~\ref{fig:9}{\it b} and {\it c}) before merging into a single protostar at $\tps \simeq16$\,yr. 
A circumstellar structure and spiral pattern then develop around the protostar.
After the merger, no further fragmentation occurs until $\tps \simeq 70$\,yr  (Figs.~\ref{fig:9}{\it d}--{\it f}). 
Thereafter, some fragmentation occurs in the outer disc region  (Fig.~\ref{fig:9}{\it g}). 
While such fragments (or protostars) are orbiting around the central massive protostar, they gradually migrate inwards  under gravitational interaction with the circumstellar disc  and finally fall onto the central protostar. 
One of these fragments evolves into a protostar before falling onto the central protostar, whereas the others fall before their density reaches the protostellar density. 
Inward migration and falling onto the central protostar are considered to typify the present-day star formation processes  \citep{vorobyov06,vorobyov10,machida10b,tsukamoto11,tsukamoto13}. 
The cloud evolution after the first protostar formation considerably differs between the $Z=10^{-3}\zsun$ and  $Z\le10^{-4}\zsun$ models; the former is more representative of present-day star formation processes.

\subsection{$Z\le10^{-2}\zsun$ Models}
The cloud evolutions and circumstellar environments in the $Z=10^{-2}$, $10^{-1}$ and $1\zsun$ models are similar to those in the $Z=10^{-3}\zsun$ model. 
Figure~\ref{fig:10} shows the time sequence of the $Z=10^{-2}\zsun$ model.
Also for this model, we find that fragments appear just before the protostar formation (Fig.~\ref{fig:10}{\it a}) and later merge into a single protostar (Fig.~\ref{fig:10}{\it b}).
The merger is followed by the formation of a stable circumstellar disc with no clear spiral structure (Figs.~\ref{fig:10}{\it c} and {\it d}). 
The inner edge of this disc fragments at $\tps \simeq 50$\,yr (Fig.~\ref{fig:10}{\it e}), but the fragment immediately falls onto the protostar and disappears (Fig.~\ref{fig:10}{\it f}). 
Although fragmentation in the circumstellar disc continues up to $\tps \simeq 70$\,yr (Fig.~\ref{fig:10}{\it g}), the fragments always move inwards and merge with the central protostar.  
At the end of the calculation, a single protostar remains, enclosed by a stable rotation-supported disc (Fig.~\ref{fig:10}{\it i}).

Figure~\ref{fig:11} shows the time sequence of the $Z=10^{-1}\zsun$ model. 
As observed in other models, the first fragmentation occurs just before the first protostar formation (Fig.~\ref{fig:11}{\it a}).
The fragments orbit each other for a while, then they merge into a single protostar. 
Although fragmentation occurs again near the protostar just after the merger (Fig.~\ref{fig:11}{\it c}), the fragment immediately merges into the protostar (Fig.~\ref{fig:11}{\it d}). 
Note that in reality, fragmentation may not occur in the very proximity of the protostar because the protostellar luminosity, which is ignored in this study, can heat the disk and suppress fragmentation \citep{stamatellos09a,stamatellos09b,romax14}. 
Thereafter, a spiral structure develops without further fragmentation, as shown in Figures~\ref{fig:11}{\it e}-{\it i}. 
At the end of the calculation, a single protostar is enclosed by a rotation-supported disc of radius of $\sim5$\,AU.

Figure~\ref{fig:12} shows the density distribution in the  $Z=1\zsun$ model at $\tps = 45.1$\,yr. 
As previously mentioned, we calculated the cloud evolution for only $55$\,yr after the protostar formation in this model.
The figure shows  a single protostar enclosed by a circumstellar disc.
The spiral pattern develops in the circumstellar disc, while no fragmentation occurs by the end of the calculation.

\subsection{Number of Fragments}
\label{sec:fragnum}
As mentioned in previous subsections, the frequencies of fragmentation and the numbers of fragments considerably differ between the models. 
Lower-metallicity clouds are prone to frequent and numerous fragmentation events, whereas higher-metallicity clouds tend to establish a single protostar enclosed by a stable rotation-supported disc. 
We quantitatively estimated the fragment numbers and protostellar masses among the models by the following procedure: 
(1) the maximum density $n_{\rm max}$ is searched for within the region of $r<20$\,AU;  
(2) when the maximum density exceeds the protostellar density $n_{\rm max} > n_{\rm ps}$, the centre of the protostar (or protostar position) is assumed to be on the cell having the maximum density;  
(3) the protostellar mass is estimated by integrating the mass over $n > 0.25 \, n_{\rm ps}$ within the region of $r_{\rm p}< 2 \, \lambda_{\rm J}$, where $r_{\rm p}$ and $\lambda_{\rm J}$ denote the radial distance from the protostellar position and the Jeans length (derived from $n_{\rm ps}$ and the corresponding temperature), respectively; 
(4) the integrating region ($n > 0.25 \, n_{\rm ps}$  within $R_{\rm ps}< 2 \, \lambda_{\rm J}$) is masked and  ignored in the next step of searching for the maximum density;
(5) steps (1)-(4) are iterated until the maximum density is less than $n_{\rm ps}$.
The number of fragments and mass of each protostar are then determined. 

This procedure was implemented at different time steps for each model. 
Note that when counting the number of protostars, we omitted the protostars that escaped from the region of $r<20$\,AU, as mentioned above. 
Note also that the density and spatial ranges over which the protostellar mass was integrated, namely $n > 0.25 \, n_{\rm ps}$ and $r_{\rm p}< 2 \, \lambda_{\rm J}$,  were decided by trial and error.
In these trials, the number of fragments output by the numerical procedure was compared with that counted manually. 
Although this procedure roughly estimated the number of protostars and their masses, it has limited applicability. 
For example,  very close binaries separated by $<2\lambda_{J}$ are recognized as a single protostar. 
However, we consider that this procedure allows a qualitative, rather than a quantitative, discussion of the relation between the fragmentation process and cloud metallicity.

Figure~\ref{fig:13} plots the number of protostars in all models against the time elapsed after the first protostar formation.
The figure clearly indicates that the number of protostars decreases as the cloud metallicity increases. 
Although we calculated the cloud evolution  for only  $\sim100$\,yr after the first protostar formation, the effect of cloud metallicity on the protostar formation process is clear.

No significant difference is apparent between the $Z=0$ and $10^{-6}\zsun$ models; both models have at most 17 or 18 protostars around the centre of the collapsing cloud. 
More protostars appear in the  $Z=10^{-5}\zsun$ model than in the $Z=0$ and $10^{-6}\zsun$ models. 
The $Z=10^{-5}\zsun$ model yields at most 25 protostars.
This difference may be attributed to the higher protostellar masses in the $Z=0$ and $10^{-6}\zsun$ models than in the $Z=10^{-5}\zsun$ model (\S\ref{sec:mass}). 
Massive stars can stabilize their circumstellar environment against gravitational instability (for details, see \S\ref{sec:stability}). 
In addition, the temperature decrease by dust cooling can affect the number of fragments, as shown in Figures~\ref{fig:6} and \ref{fig:7b}.
Considerably fewer protostars develop in the $Z=10^{-4}\zsun$ model than in the $Z<10^{-4}\zsun$ models. 
Although 10 protostars appear in the $Z=10^{-4}\zsun$ model at $\tps \sim80$\,yr, most of them subsequently merge into a high-mass protostar, yielding a single star at $\tps\sim90$\,yr. 
However, disc fragmentation yields  several protostars around the most massive protostar for $\tps\gtrsim100$\,yr.

Fragmentation is rare in the $Z>10^{-4}\zsun$ models.
Although two protostars are formed in the very early accretion phase ($\tps\lesssim 10$\,yr) in the $Z=10^{-3}$, $10^{-2}$ and $10^{-1}\zsun$ models, they merge into a single protostar. 
A rotation-supported circumstellar disc then develops around the merged protostar.
Although the disc occasionally fragments, the fragments (or protostars) eventually fall onto the central protostar. 
Consequently, these models always yield a single protostar at the end of the calculation.
The $Z=1\zsun$ model also produces a single protostar with no fragmentation. 
In the $Z>10^{-4}\zsun$ models, the spiral pattern induced in the disc appears to facilitate outward transfer of angular momentum. 
In addition, fragmentation is likely suppressed by the lower mass accretion rate during the main accretion phase.

\subsection{Protostellar Mass}
\label{sec:mass}
Figure~\ref{fig:14} plots the mass of the most massive protostar in each model against the time elapsed after the first protostar formation. 
Note that this figure does not necessarily trace the same protostar because the most massive protostar among multiple protostars can vary from one epoch to another.
Note also that since we integrated the high-density gas in a region (\S\ref{sec:fragnum}) without a sink, the protostellar mass may be somewhat overestimated or underestimated. 
In this figure, a sudden increase indicates merging of a less massive protostar, while a sudden decrease can result from  fission or tidal gas stripping by another protostar \citep{machida13a}.  
The sudden increase and decrease are also caused by changes in the corresponding protostar. 
At $\tps \sim 100$\,yr, the most massive protostar is $\sim 1\msun$ and $\sim 0.05\msun$ in the $Z\le10^{-4}\zsun$ and  $Z>10^{-4}\zsun$ models, respectively. 
Thus, the mass is accreted at an estimated rate of $\sim 0.01\msun$\,yr$^{-1}$ in the $Z\le10^{-4}\zsun$ models and $\sim5\times10^{-4}\msun$\,yr$^{-1}$ in the $Z>10^{-4}\zsun$ models. 
The mass accretion rates in different classes ($Z\le10^{-4}\zsun$ and $Z>10^{-4}\zsun$) differ by a factor of $\lesssim10$ and roughly agree with those in Figure~\ref{fig:2}. 
To understand the protostellar evolution, we need to calculate a further evolutionary stage because the protostellar mass is much less than the host cloud mass even at the end of the calculation.

Figure~\ref{fig:15} shows the protostellar masses in the $Z=0$, $10^{-6}$, $10^{-5}$, $10^{-4}$, $10^{-3}$ and $10^{-2}\zsun$ models, in which the masses of the first, second, fourth, sixth, eighth and tenth most massive protostars are plotted against the time elapsed after the first protostar formation. 
The total mass of the protostars is also plotted by a broken line in each panel. 
Since the number of protostars is less than $N_{\rm ps}=10$ in the $Z>10^{-4}\zsun$ models, fewer than seven lines are plotted in  the panels of $Z=10^{-3}$ and $10^{-2}\zsun$. 
This figure indicates that the masses of the most and second-most massive protostars are very similar in the $Z=0$, $10^{-6}$ and $10^{-5}\zsun$ models. 
In the $Z=10^{-4}\zsun$ model, however, the most massive protostar is substantially more massive than the others.
Moreover, in the $Z=10^{-3}$ and $10^{-2}\zsun$ models, the secondary protostar is considerably less massive than the most massive protostar for $\tps \gtrsim 60$\,yr.
Note that in these models, the masses of the most massive and secondary protostars are comparable only when fragmentation occurs near the primary protostar.

In the $Z\le 10^{-4}\zsun$ models, fragmentation occurs everywhere in the cloud, and no stable rotation-supported disc is formed.  
Therefore, although the protostars interact with each other, they essentially evolve independently. 
Consequently, several protostars of similar mass appear around the centre of the collapsing cloud.
On the other hand, the $Z>10^{-4}\zsun$ models yield a single protostar enclosed by a stable rotation-supported disc, which occasionally fragments.  
The disc develops a spiral structure that effectively transfers the angular momentum outwards, suppressing frequent fragmentation and new protostar formation.
In addition, if a fragment or protostar forms in the disc, it eventually falls onto the central protostar.
Thus, only the central protostar alone can significantly increase its mass in the $Z>10^{-4}\zsun$ models.

\section{Discussion}
\label{sec:discussion}
\subsection{Star Formation Mode}
\label{sec:mode}
The calculation results indicate that the star formation mode considerably differs in the lower- and higher-metallicity clouds. 
In the lower-metallicity clouds ($Z\le10^{-4}$), no stable disc is formed, and frequent fragmentation leads to abundant protostars. 
Although some protostars are ejected from the central region, the remainder form a small cluster of $\sim10-20$ protostars, as shown in Figures~\ref{fig:6}-$\ref{fig:8}$. 
The star formation process in the $Z=10^{-6}$, $10^{-5}$ and $10^{-4}\zsun$ models is very similar to that in the primordial ($Z=0$) case \citep{clark08, clark11a, clark11b, smith11,greif11,greif12,machida13a}.
On the other hand, the higher-metallicity clouds ($Z > 10^{-4}\zsun$ models) yield  a single protostar enclosed by a circumstellar disc, as shown in Figures~\ref{fig:9}-\ref{fig:12}.
Although fragmentation occasionally occurs in the disc, each fragment eventually falls onto the central protostar. 
The star formation process in the $Z=10^{-3}$, $10^{-2}$ and $10^{-1}\zsun$ models is similar to the present-day case {\citep{vorobyov06,vorobyov10,machida10b,tsukamoto11,tsukamoto13}, in which fragmentation rarely occurs in the disc and the fragments usually fall onto the central protostar. 
The different star formation modes between clouds of lower-metallicity (small stellar cluster) and higher-metallicity (star-disc system) are presumably due to differences in both the lifetime of the first core and  mass accretion rate.

\subsubsection{Effects of First Hydrostatic Core}
\label{sec:first-core}
We first discuss the effects of the first core on the star formation process. 
In present-day star formation (or the solar-metallicity case), after the collapsing gas becomes optically thick to dust thermal emission at $n\sim10^{10}\cm$, the gas behaves adiabatically and the first (hydrostatic) core forms in the collapsing cloud \citep{larson69}.  
The size and mass of the first core is $\gtrsim0.01$\,AU and $\gtrsim0.01\msun$, respectively \citep{masunaga00}, depending on the rotation rate of the star-forming cloud \citep{saigo06}.
As the gas accretes onto the first core, its central density gradually increases and the second collapse, which is induced by dissociation of the molecular hydrogen, occurs {\em only} in the central part of the first core at $n_{\rm c}\sim 10^{15-16}\cm$. 
The gas again becomes adiabatic and the protostar (or second hydrostatic core) of radius of $\sim \rsun$ and mass of $\sim 10^{-3}\msun$, finally appears. 
The thermal evolution of a non-rotating present-day cloud is also revealed in Figure~\ref{fig:1}.

Since the rotation of the molecular cloud is non-negligible \citep{goodman93,caselli02}, the first core is partly supported by the rotation (the thermal pressure is the primary support). 
With the rotation, the first core develops a spiral structure that effectively transfers the angular moment outwards, suppressing  fragmentation before the protostar formation \citep{bate98,machida08a,tomida13}. 
After the protostar formation, the first core remains and evolves into a circumstellar (or rotation-supported) disc. 
Only the central part of the first core alone becomes the protostar, whereas the remainder rotates around the protostar.
Thus, the protostar at its formation epoch is already enclosed by a large-scale disc-like structure evolved from  the first core or its remnant \citep{bate98,bate11,machida10a,machida12}. 
Since the remnant of the first core or circumstellar disc is more massive than the protostar just after the protostar formation, fragmentation can occasionally occur in the disc \citep{inutsuka10,inutsuka12}. 
However, because the fragments usually fall onto the first formed protostar,  a single protostar remains in the early main accretion phase  \citep{vorobyov06,walch09, vorobyov10,machida11a,tsukamoto11,walch12,tsukamoto13}.
The present-day star formation process has been well investigated, and the above scenario is widely accepted.
In addition, some first core candidates have been confirmed by observations \citep{chen10,enoch10,dunham11,belloche11,pineda11,chen12,pezzuto12,tsitali13,hirano14b}.

On the other hand, the lack of dust prevents the first core formation in the primordial collapsing cloud. 
As seen in Figure~\ref{fig:1}, the gas temperature gradually increases in the range of $10^4\cm \lesssim n \lesssim 10^{20}\cm$ with a polytropic index of $\gamma\sim1.1$ in the primordial ($Z=0$) clouds, but suddenly increases at $n\sim 10^{10}\cm$ in the present-day ($Z=\zsun$) clouds. 
Consequently, in the primordial case,  no significant structure develops during the protostar formation epoch (or gas-collapsing phase). 
In addition, clouds with solar metallicity virtually cease collapsing when the first core forms.
The rotational timescale then becomes shorter than the collapsing timescale, and rotational motion dominates in the first core.
On the other hand, in primordial clouds, where the collapsing timescale is always shorter than the rotational timescale with $\gamma \sim 1.1$, the gas continues to fall towards the cloud centre or onto the  protostar \citep{saigo00}. 
Note that although fragmentation can occur prior to the protostar formation if the primordial star-forming cloud has a sufficiently high angular momentum, the gas collapse continues in each fragment \citep{machida08}.
Consequently, the structure around the protostar during the protostar formation epoch considerably differs between the present-day and primordial clouds. 
The protostar is enclosed by a disc in present-day clouds, but is surrounded by radially infalling gas in primordial clouds.
The difference in circumstellar environment during the protostar formation epoch influences further evolution of the protostar and circumstellar region.

The density at which the first core forms depends strongly on the metallicity $Z$, as seen in Figure~\ref{fig:1} (see also Figs.~2 and 3 in \citealt{machida09a}). 
The first core appears when the cloud has a metallicity of $Z \gtrsim 10^{-5}\zsun$. 
The size and mass of the first core decrease as the cloud metallicity lowers \citep{machida09a}.
To investigate the effect of the first core, we present the cloud evolution prior to the protostar formation in Figure~\ref{fig:16}. 
We observed that the first core is clearly established in the  $Z=10^{-2}$ and $10^{-3}\zsun$ models.
At its formation epoch,  the size of the first core is  $\sim5$\,AU in the $Z=10^{-2}\zsun$ model and $\sim3$\,AU in the $Z=10^{-3}\zsun$ model. 
In both models, the first core increases in size as the gas accretes on it, and a spiral structure develops. 
Moreover, fragmentation occurs before the cloud maximum density reaches the protostellar density in the $Z=10^{-3}\zsun$  model. 
Consequently, in both $Z=10^{-3}$ and $10^{-2}\zsun$ models, the protostar at its formation epoch is enclosed by a rotating disc or spiral structure.
The sizes of these structures are $\sim20$\,AU ($10^{-2}\zsun$ model) and $\sim5$\,AU ($10^{-3}\zsun$ model), respectively.

As seen in Figure~\ref{fig:16}, the gas in the $Z=10^{-4}\zsun$ model continues to collapse for $\tps \lesssim -5$\,yr, and  the first core appears just before the protostar formation at $\tps\simeq -5$\,yr. 
The first core forms at high gas density of $\sim10^{16}\cm$, and is as small as $\lesssim1$\,AU.  
The first core transforms into a ring-like structure that fragments into two protostars.
In the $Z=10^{-5}\zsun$ model, no clear first core appears until the protostar formation. 
Thus, during the formation epoch, the protostar has no disc-like or spiral structure around it. 
Evidently, protostars in clouds of higher metallicity ($Z>10^{-4}\zsun$) and low-metallicity ($Z\le10^{-4}\zsun$) form and evolve in rather different circumstellar environments.
It appears that the first core does not significantly affect the main accretion phase in clouds with metallicity of $Z\le10^{-4}\zsun$.
According to Figures~\ref{fig:6}-\ref{fig:12}, the star formation process changes between $Z=10^{-3}$ and $10^{-4}\zsun$.
Fragmentation is frequent and leads to many protostars with no stable disc formation in the $Z\le10^{-4}\zsun$ models, whereas the $Z > 10^{-4}\zsun$ models generate a single protostar enclosed by a stable rotation-supported disc.
This difference appears to be related to  the existence of the first core.

\subsubsection{Mass Accretion Rate and Disc Stability}
\label{sec:stability} 
In this subsection, we analytically discuss the relation between the disk stability and mass accretion rate.
The star formation process is expected to depend also on the mass accretion rates. 
As reported in previous studies \citep[e.g.][]{hosokawa09b,dopcke13}, the mass accretion rate onto the circumstellar region decreases as the cloud metallicity increases.
This phenomenon is attributed to the cloud temperature.
The metal-rich cloud has a lower temperature and the mass accretion rate is roughly proportional to $T_{\rm cl}^{3/2}$, where $T_{\rm cl}$ is the temperature of the initial star-forming cloud \citep{larson03}.
As shown in \S\ref{sec:results}, fragmentation frequently occurs around the cloud centre in the lower-metallicity clouds, whereas higher-metallicity clouds rarely fragment. 
The stability of the circumstellar disc evolving around the protostar is considered to be determined by Toomre's Q parameter $ Q \equiv (c_s \kappa)/(\pi G \Sigma)$,
where $c_s$, $\kappa$ and $\Sigma$ are the sound speed, epicyclic frequency and surface density, respectively \citep{toomre64}.
The epicyclic frequency $\kappa$ can be replaced by the Kepler angular velocity $\Omega_{\rm K}$ [$\equiv (GM_{\rm ps}/r^3)^{1/2}$] when Kepler rotation is realized in the disc. 
For simplicity, we discuss the disc stability in terms of the parameter 
\begin{equation}
Q\equiv \dfrac{c_s \Omega_{\rm K}}{\pi G \Sigma}, 
\label{eq:toomre}
\end{equation}
and  fragmentation is considered to occur when $Q<1$.
In our calculation, since some models develop no clear rotation-supported disc, we cannot assess the adequacy of Toomre's analysis.
We tentatively use it to roughly analyse the stability around the protostar.
To simplify the problem, we also assume constant sound speed. 
This assumption is applicable because we investigate the evolution of the circumstellar environment over a sufficiently short time ($\sim100$\,yr after the first protostar formation).
During this early phase, protostellar heating can be ignored because the environmental temperature is initially high near the protostar \citep{omukai10}.

Equation~(\ref{eq:toomre}) implies that the central star tends to stabilize the disc on a Keplerian timescale $t_{\rm Kep}=2\pi/\Omega_{\rm K}$, while the increasing disc surface density tends to destabilize the disc on a disc growth timescale $t_{\rm grow}=M_{\rm disc}/\dot{M}$.
Here, $M_{\rm disk}$ and $\dot{M}$ are the disc mass and mass accretion rate onto the central region, respectively. 
During the main accretion phase, the disc growth timescale may be shorter than the Keplerian timescale. 
In such a case, $Q$ continues to decrease and fragmentation is expected to be induced.
To very roughly investigate the stability of the disc, we compared the Keplerian timescale $t_{\rm Kep}$ with the disc growth timescale $t_{\rm grow}$. 
Figure~\ref{fig:17} plots these timescales in the $Z=10^{-5}$, $10^{-4}$, $10^{-3}$ and $10^{-2}\zsun$ models for different protostellar masses, namely  $M_{\rm ps}=1$ (all models), $0.1$ ($Z=10^{-4}$, $10^{-3}$ and $10^{-2}\zsun$ models) and $10\msun$ ($Z=10^{-5}\zsun$ model).
The mass accretion rate is estimated with a sink (see \S\ref{sec:psmodel} and Fig.~\ref{fig:2}), and the disc mass is assumed as 1\% of the central protostellar mass, that is, $M_{\rm disc}=0.01\,M_{\rm ps}$. 
The protostellar radii $R_{\rm ps}$ at each epoch (Fig.~\ref{fig:3}) are also plotted in this figure.
Since the region of $r<R_{\rm ps}$ defines the inner region of the protostar, it is excluded from the disc stability analysis  (we call this region the forbidden region).

Outside the protostar ($r>R_{\rm ps}$), we assume that the disc is stable when $t_{\rm Kep} < t_{\rm grow}$ and  unstable when $ t_{\rm Kep} > t_{\rm grow}$. 
In the latter case, it is expected  that the disc fragments before the central star restores the disc surface density to a stable state. 
In other words, the denominator of equation~(\ref{eq:toomre}) rapidly increases and $Q < 1$ is soon realized when $ t_{\rm Kep} > t_{\rm grow}$. 
Figure~\ref{fig:17} indicates that when $M_{\rm ps}=1\msun$ (red lines), the whole region is gravitationally unstable in the $Z=10^{-5}\zsun$ model. 
In the $Z=10^{-4} \zsun$ model,  a small stable region exists in the range of $ 0.25 \,{\rm AU} \lesssim r \lesssim 1.5\,{\rm AU}$. 
In the $Z=10^{-3}$ and $10^{-2}\zsun$ models, the stable region enlarges to $0.05\,{\rm AU} \lesssim r \lesssim 10-30\, {\rm AU}$.
Thus, the stable region expands as the metallicity increases.
Moreover, in the $Z=10^{-5}\zsun$ model, the stable region appears only when $M_{\rm ps}>1 \msun$. 
This very rough estimate (there are many ambiguous  parameters) suggests that fragmentation occurs without stable disc formation in the lower-metallicity models because the disc is destabilized by the high mass accretion rate before the disc reaches a stable state.
On the other hand, Figure~\ref{fig:17} indicates that a stable circumstellar disc is maintained in the  $Z>10^{-4}\zsun$ models even when the protostellar mass is as small as $0.1\msun$ (black lines).

Figure~\ref{fig:18} shows the density and velocity distributions at $\tps\simeq100$\,yr in the $Z=10^{-5}$, $10^{-4}$, $10^{-3}$ and $10^{-2}\zsun$ models. 
At this epoch, the most massive protostars are $1.5\msun$ ($Z=10^{-5}\zsun$), $1\msun$ ($Z=10^{-4}\zsun$), $0.04\msun$ ($Z=10^{-3}\zsun$) and $0.04\msun$ ($Z=10^{-4}\zsun$).
The $10^{-5}\zsun$ and $10^{-4}\zsun$ models develop no stable disc, and many fragments form around the cloud centre; in contrast, a disc-like structure is confirmed in the $10^{-3}$ and $10^{-2}\zsun$ models. 
Thus, the simulation results approximately agree with our simple estimate shown in Figure~\ref{fig:17}.
In summary, a high mass accretion rate leads to vigorous fragmentation, whereas a lower rate permits a stable rotation-supported disc formation. 
As the protostellar mass increases, a stable disc should form even in the lower-metallicity clouds. 
However, at least during the very early phase of the star formation, fragmentation frequently occurs and many protostars form around the centre of the collapsing cloud.
Although further calculations are necessary to determine the fate of protostars, the protostellar environments of lower- and higher-metallicity clouds are considerably different.

\clearpage
\subsubsection{Gas Temperature and Disk Stability}
Although the $Z=0$, $10^{-6}$ and $10^{-5}\zsun$ models have almost the same accretion rate (Fig.~\ref{fig:2}), the number of fragments is larger in the $Z=10^{-5}\zsun$ model than in the $Z=0$ and $10^{-6}\zsun$ models (Fig.~\ref{fig:13}). 
This is expected to be due to the difference in thermal evolution.

As shown in Figure~\ref{fig:6}, fragmentation tends to occur in a lower-temperature region in the $Z=10^{-6}\zsun$ model.
On the other hand, in the $Z=10^{-5}\zsun$ model, the cloud centre has a relatively low temperature and the entire fragmentation region is embedded in the low-temperature region, as shown in Figure~\ref{fig:7b}.
The temperature decrease in the high-density region ($10^{12} \cm \lesssim n \lesssim 10^{16}\cm$ of Fig.~\ref{fig:1}) is caused by dust cooling \citep{omukai10}. 
Figures~\ref{fig:1}, \ref{fig:6} and \ref{fig:7b} indicate that the gas temperature around the protostar (or fragmentation region) is lower in the $Z=10^{-5}\zsun$ model than in the $Z=0$ and $10^{-6}\zsun$ models.
The lower temperature and smaller sound speed decrease the $Q$ parameter and make the disk unstable.
A temperature gap in the high-density region is also seen in the $Z>10^{-5}\zsun$ models.
However, such models have a relatively small mass accretion rate (Fig.~\ref{fig:2}). 
Both a considerably large accretion rate and lower-temperature environment are realized in model $Z=10^{-5}\zsun$.
Thus, it is expected that the disk in the $Z=10^{-5}\zsun$ model is the most unstable and many fragments will appear.
A similar result was presented in \citet{tanaka14}.

\subsection{Effects of Initial Conditions}
\label{sec:init}
In this study, as the initial state, we adopted the Bonner-Ebert sphere, which is in a near-equilibrium state.
The properties of Bonner-Ebert clouds are determined by the central density and temperature, and the cloud temperature is determined by the cloud metallicity (see \S\ref{sec:model}).
Therefore, the cloud masses and sizes differ between models, as listed in Table~\ref{table:1}. 
Instead, as in previous studies \citep{dopcke11,dopcke13,safranek14}, even when the star-forming clouds have different metallicities, we can set an identical cloud as the initial condition in which the initial cloud has the same mass and size independent of its metallicity. 
However, in such a setup, the cloud stabilities differ between models because the initial cloud is in a non-equilibrium state (the equilibrium state is determined by the cloud temperature or metallicity). 
In some previous studies, although the (isothermal) temperature was assumed to be identical in initial clouds with different metallicities, the cloud temperature would rapidly change and each cloud would have different thermal energies (or different stability) just after the cloud began to collapse.

The (initial) cloud stability, defined as the ratio of thermal to gravitational energy $\alpha_0$, significantly affects the fragmentation condition in the collapsing cloud.
Clouds with lower thermal energies (smaller $\alpha_0$) are prone to fragmentation \citep{tsuribe99a, tsuribe99b}. 
Thus, it is expected that under identical initial conditions, a higher-metallicity cloud will more likely fragment because the thermal energy is reduced by efficient cooling. 
In this scenario, the initial condition determines the outcome (or frequency of fragmentation). 
Especially, unless the high-density gas region or the first core is not resolved with a sink, frequent fragmentation is expected in higher-metallicity clouds because $\alpha_0$ is initially small, as seen in  previous studies.

It is very difficult to determine adequate initial conditions for the general star formation process. 
The initial conditions of a prestellar cloud with zero metallicity may be determined by cosmological simulations \citep[e.g.][]{hirano14}, whereas those of a solar-metallicity cloud are deduced from observations \citep[e.g.][]{motte98}.
In both cases,  initial clouds are considered to be well described by the Bonner-Ebert density profile (see \citealt{abel02} for the primordial case and \citealt{alves01} for the solar metallicity case). 
However, the properties of prestellar clouds (or adequate initial conditions) with  $0 < Z < \zsun$ cannot be determined. 
Although it seems natural to assume that such clouds also exist in a near-equilibrium state with a Bonnor-Ebert density profile, their properties cannot be confirmed.

The fragmentation condition depends not only on the thermal energy but also on the initial cloud rotation. 
Clouds with initially higher rotational energy (or larger $\beta_0$) are more likely to fragment. 
In our previous studies, we investigated the effects of the initial rotation on the star formation process in both primordial \citep{machida08c,machida13a} and solar-metallicity \citep{machida05b,machida08, machida10a,machida12} clouds. 
These studies showed that, independent of $\beta_0$, solar-metallicity clouds develop a stable rotation-supported disc,  whereas primordial clouds undergo vigorous fragmentation, leading to numerous protostars with no stable discs. 
Identical results were obtained in the primordial and solar-metallicity clouds investigated in the present study. 
To better understand the accretion phase of star formation in clouds with different metallicities, we may need to investigate the cloud evolution with different $\beta_0$. 
However, such a parameter survey is currently precluded by the  very long calculation time required  to investigate the accretion phase resolving the protostar.
The typical computational time of a single model in this study was six months. 
Although we expect that changing $\beta_0$ will not qualitatively change the cloud evolution, as previously found in primordial clouds  \citep{machida13a} and solar-metallicity clouds with a sink \citep[][]{machida10a},  the effect of rotation in clouds with different metallicities requires further investigation.

In this study, we limited the number of models and focused on the effect of thermal evolution, which is dependent on cloud metallicity.
Therefore, we constrained the initial cloud  to a near-equilibrium condition and a constant ratio of rotational to gravitational energy. 
However, determining the correct initial conditions is essential for understanding the evolution of clouds with different metallicities. 
Thus, in future work, we must investigate cloud evolution over a large parameter space of $\alpha_0$ and $\beta_0$.

\subsection{Effects of Magnetic Field}
\label{sec:mag}
In this study, we ignored the effects of the magnetic field.
However, the magnetic field has a large impact on star formation, because it effectively transfers the angular momentum, thereby suppressing  cloud fragmentation if sufficiently strong \citep{machida08a,hennebelle08}.
In addition, a portion of the infalling gas is ejected by the protostellar outflow driven by the Lorentz force \citep{tomisaka02,machida08b}.
Disc formation is also suppressed by magnetic braking \citep{mellon08,mellon09,machida13a}. 

To investigate the effects of the magnetic field in clouds with different metallicities, we require both the magnetic field strength of the star-forming cloud and the dissipation process of the magnetic field in each cloud (or each metallicity).
The magnetic field strength of solar-metallicity clouds is determined from observations of nearby star-forming regions \citep{crutcher99}.
On the other hand, the field strengths of clouds with $Z\ll \zsun$ are less easily determined  because there is no observational evidence.

In addition, the magnetic field can dissipate and weaken in a collapsing cloud. 
In clouds with solar metallicity, the magnetic field is significantly reduced by Ohmic dissipation \citep{nakano02,machida07}. 
On the other hand, collapsing primordial clouds do not lose their magnetic fields by dissipation \citep{maki04,maki07}. 
The difference in the magnetic dissipation process between solar-metallicity and primordial clouds is primarily due to the different ionization degree. 
Clouds with solar metallicity are relatively poorly ionized because of their lower temperature and abundant dust.
In contrast, a primordial gas has a high temperature and lack of dust, and is ionized to a relatively high degree.  
Since the temperature of the collapsing cloud and dust abundance depend on the metallicity, the dissipation degree of the magnetic field should also be affected by the metallicity.  
Although the initial strength and dissipation degree of the magnetic field are not easily determined, investigating the effects of  the magnetic field on star formation processes in clouds of different metallicities is an important future task. 

\subsection{Effect of Stellar Feedback}
In this study, we ignored the effect of  stellar feedback that can suppress disk fragmentation through heating of the circumstellar disk by the protostellar luminosity and  stabilization against gravitational instability.
In this subsection, we discuss the effect of protostellar feedback on fragmentation.

In present-day low-mass star formation, the protostellar feedback affects the disk evolution and fragmentation process. 
\citet{offner09} investigated low-mass star formation in a turbulent  cloud and showed that protostellar feedback tends to suppress disk fragmentation \citep[see also][]{bate09,krumholz10}. 
\citet{stamatellos12} showed that episodic accretion alleviates the suppression of disk fragmentation due to  protostellar feedback because fragmentation can occur in a quiescent phase between outbursts \citep{lomax14}.
Thus, in the present-day star formation process, although protostellar feedback cannot completely suppress disk fragmentation, it plays an important role for the disk evolution. 

On the other hand, the effect of stellar feedback is not very significant in primordial and lower-metallicity clouds.
\citet{stacy12} showed that radiative feedback does not prevent disk fragmentation and multiple stars appear in primordial clouds.
\citet{smith11} calculated the evolution of primordial minihalos including the luminosity from accreting protostars, and showed that accretion luminosity does not prevent disk fragmentation. 
This is because the heating from the accretion luminosity contributes little to the disk thermal evolution; the compression heating dominates other heating processes. 
To estimate the effect of protostellar feedback in clouds with different metallicities, \citet{omukai10} compared the cloud gas temperature before heating (or before the protostar formation) with that heated by the protostar, and showed that the former dominates the latter in clouds with $Z\le 10^{-3}\zsun$. 
In lower-metallicity clouds, since the gas temperature is intrinsically high due to the deficit of efficient coolant, protostellar feedback does not play a significant role for disk evolution and fragmentation.
On the other hand, in higher-metallicity clouds, since the gas temperature is relatively low, the protostellar luminosity effectively heats the disk and suppresses fragmentation.

In our study, fragmentation frequently occurs in lower-metallicity clouds, while it rarely occurs in higher-metallicity clouds. 
Since previous studies indicate that the effect of protostellar feedback is not significant in lower-metallicity clouds, we expect that frequent fragmentation occurs in lower-metallicity clouds, as seen in primordial clouds \citep{greif12,stacy14}.
In addition, at the end of the calculation, the protostars are found to have masses of $\sim1\msun$ in lower-metallicity clouds ($Z \le 10^{-4}\zsun$) and $<0.1\msun$ in higher-metallicity clouds ($Z> 10^{-4}\zsun$).
Thus, it is expected that the protostars exhibit minimal radiative effects during the calculation.
However, to quantitatively understand fragmentation properties such as stellar mass and the number of fragments, we need to calculate the cloud evolution with stellar feedback.

\subsection{Can Ejected Low-metal Stars Be Observed?}
\label{sec:obs} 
In clouds with $Z\le 10^{-4}\zsun$, many protostars form by fragmentation and  low-mass protostars are preferentially ejected from the cloud centre.
In addition, low-mass protostars with masses of $\ll 1\msun$  exist even around the centre of the cloud at the end of the calculation.
These protostars may evolve into zero- (or extremely low-) metallicity ($Z\le10^{-4}\zsun$) and low-mass ($M\lesssim 0.8\msun$) stars that can survive until the present day and may be observable.
In this subsection, we discuss the observation probability of such zero-metallicity or extremely metal poor stars in our galaxy. 

We evolved each cloud $\sim 100$\,yr after the first protostar formation and found that fragmentation frequently occurs and many protostars appear in zero- and lower-metallicity environments.  
Some protostars have a mass of $\ll 1\msun$ at the end of the calculation.
However, it is expected that such low-mass protostars increase their mass to reach $>1\msun$ in a further evolution stage because the zero- (or low-) metallicity host cloud has a sufficient mass to sustain the mass accretion for a long duration.
In addition, the mass accretion rate is considerably high in zero- and lower-metallicity clouds.
Thus, we expect that it is considerably difficult for a protostar around the cloud centre to keep its mass within $\lesssim 0.8\msun$. 
The stars with a mass of $>0.8\msun$  would have died.

On the other hand, since the ejected protostar moves away from the cloud centre and mass accretion is expected to gradually weaken, such stars may have a mass of  $\lesssim 0.8\msun$.
Although we need to calculate the cloud in further evolution stages to determine the final stellar mass and the number of zero-metallicity (or extremely metal poor) stars still surviving, we probatively estimate the number of observable zero-metallicity stars in our galaxy. 
We define the number of low-mass stars with a mass of $\le0.8\msun$ in a single minihalo as $N_{\rm mh}$.
The minihalo is assumed to have a baryonic mass of $M_{\rm mh}=10^6\msun$, while the mass of our galaxy is $M_{\rm gal}\sim10^{11}\msun$.
The galaxy evolves by the merger of minihalos and we simply assume that about $10^5$ minihalos are gathered to form our galaxy. 
Thus, our galaxy contains $N\times10^5$ zero-metallicity stars.
Therefore, the number ratio of zero-metallicity  to Population I and II stars  is $N \times 10^{-6}$. 
Simply assuming $N=1$ implies that we can find a zero-metallicity star randomly in the observation of one million stars in our galaxy.
In other words, we need to observe about $10^6$ stars in our galaxy to find a zero-metallicity star. 
Although several hundred thousand stars have been spectroscopically identified \citep[e.g.][]{yanny09}, further observations may be necessary to find zero-metallicity stars.

Although the observational probability of zero-metallicity stars strongly depends on $N$,  we expect that $N$ is not large.
Since the ejected protostar can acquire its mass from the infalling envelope even after being ejected from the central region, it is difficult to keep its mass in the range of $<0.8\msun$. 
In addition, some low-mass protostars are expected to be further ejected from minihalos and may drift in intergalactic space. 
Moreover, an initially zero-metallicity star can be polluted by a binary companion \citep[e.g.][]{suda04} or interstellar medium \citep[e.g.][]{komiya10,johnson14},  and may be observed as an extremely metal poor star.
Although a more detailed study is necessary to determine the observational probability of zero- (and extremely low-) metallicity stars, we may observe them in the near future.

\section{Summary}
In this study, we investigated the early main accretion phase of star formation in clouds with different metallicities in the range of $0\le Z \le \zsun$. 
We calculated the cloud evolution by a three-dimensional nested grid code, in which the thermal evolution given by the one-zone model was used.
As the initial state, we adopted the Bonner-Ebert cloud, which exists in a near-equilibrium state. 
Each initial cloud had different metallicities and temperatures.
In other words, the initial clouds differed in mass and size, but were assigned the same ratio of thermal ($\alpha_0$) and rotational ($\beta_0$) energy to gravitational energy.   
Starting from the prestellar stage, we calculated the cloud evolution for $\sim100$\,yr after the first protostar formation, resolving the protostellar radius without introducing a sink. 
To adequately calculate the circumstellar environment, we adopted a protostellar model, which relates the protostellar radius to the protostellar mass. 

We showed that the star formation process considerably differs between the lower- and higher-metallicity clouds.
In the lower-metallicity clouds ($Z\le10^{-4}\zsun$), fragmentation frequently occurs without a stable disc formation and over $\sim10$ protostars appear at the same epoch around the cloud centre. 
The number of protostars is slightly smaller in the $Z=10^{-4}\zsun$ model than in the $Z<10^{-4}\zsun$ models.
Some of the protostars fall onto a more massive protostar. 
Thus, a massive protostar acquires its mass by two processes: gas accretion and the merger of less massive protostars.
The most massive protostar has a mass of $\sim1\msun$  at the end of the calculation.
On the other hand, less massive protostars with masses of $\sim0.01\msun$ are preferentially ejected from the central region by mutual gravitational interactions. 
They are expected to become lower-metallicity low-mass stars. 
At the end of the calculation, the cloud centre is surrounded by 7-20 protostars forming a small stellar cluster. 

On the other hand, in the higher-metallicity clouds of $Z>10^{-4}\zsun$, fragmentation rarely occurs and a single protostar appears. 
Immediately after its formation, the protostar is enclosed by a circumstellar disc, which is supported by rotation. 
Since the disc and protostar have similar masses, the disc fragments and develops a few clumps.
The clumps fall onto the central protostar before evolving into separate protostars. 
The disc develops spiral arms, which effectively transfer the angular momentum outwards. 
At the end of the calculation, in these models, a single protostar is enclosed by a circumstellar disc, which evokes the classical star formation process.  

In summary, the $Z\le10^{-4}\zsun$ models yields a small stellar cluster with no stable disc, whereas $Z>10^{-4}\zsun$ produces a protostar surrounded by a circumstellar disc.
In other words, lower-metallicity clouds fragment everywhere, and protostars independently evolve with mutual interaction but no stable disc formation.
Their higher-metallicity counterparts yield a single protostar enclosed by a large-scale circumstellar disc, which only occasionally fragments.
This evolutionary difference is caused by metallicity-dependent thermal evolution and the mass accretion rate.
In the $Z>10^{-4}\zsun$ models, the protostar formation is proceeded by a clear first core formation, and the first core ultimately becomes the circumstellar disc. 
The first core (or circumstellar disc) stabilizes the circumstellar environment against fragmentation with effective outward transfer of angular momentum.
This scenario is consistent with the present-day (or solar metallicity) star formation process.
On the other hand, no (clear) first core forms in the  $Z\le10^{-4}\zsun$ models.
Thus, the gas directly falls onto or near the protostar, and fragmentation frequently occurs without an effective mechanism of angular momentum transfer. 
Therefore, a stellar cluster composed of $\sim10$ protostars appears around the cloud centre.
This scenario mirrors the primordial star formation process.
Furthermore,  the high mass accretion rate prevents stable disc formation in the $Z\le10^{-4}\zsun$ models.
The mass accretion rate is moderated in the $Z > 10^{-4}\zsun$ models, permitting the formation of a stable circumstellar disc.

In this study, we showed that the star formation process transition occurs at $Z=10^{-4}-10^{-3}\zsun$.
However, we calculated the cloud evolution over a very limited time: $\sim 100$\,yr after the first protostar formation. 
We also ignored the magnetic field, which is expected to have a large impact on star formation.
To properly clarify the effect of metal abundance on the star formation process, we must calculate the cloud evolution over a longer time and include the magnetic field and its dissipation process by imposing adequate initial conditions. 

\section*{Acknowledgments}
The authors thank K.~Omukai and  K.~Doi for providing the one-zone thermal evolution calculation code.
The authors also thank T.~Hosokawa for providing the mass radius relation of protostars calculated by his stellar evolution code. 
We have benefited greatly from discussions with H.~Susa.
We are very grateful to an anonymous reviewer for a number of very useful suggestions and comments.
This research used computational resources of the HPCI system provided by NEC SX-9 at Cybermedia Center Osaka University and Cyberscience Center Tohoku University through the HPCI System Research Project (Project ID: hp140065).
This work was supported by JSPS KAKENHI Grant Numbers 25400232, 26103707.


\clearpage
\begin{table}
\caption{Models simulated in this study and their results.
Column 1 gives the model name. 
Column 2 gives the cloud metallicity ($Z$). 
Columns 3-5 give the temperature ($T_{\rm cl}$), mass ($M_{\rm cl}$) and radius ($R_{\rm cl}$) of the initial cloud.
Column 6 gives the protostellar density ($n_{\rm ps}$).
Column 7 gives the cell width of the finest grid.
}
\label{table:1}
\begin{center}
\begin{tabular}{c|cccccc|cccc} \hline
{\footnotesize Model} & 
$Z$ & $T_{\rm cl}$ [K] & $M_{\rm cl}$ [$\msun$] & $R_{\rm cl}$ [AU] & $n_{\rm ps}$ [$\cm$] & $h$ [AU] \\
\hline
1 & 0              &197 & 2500  & $3.8\times10^5$ & $3.4\times10^{17}$ & $3.8\times10^{-2}$ \\
2 & $10^{-6}\zsun$ &195 & 2460  & $3.8\times10^5$ & $3.8\times10^{17}$ & $3.8\times10^{-2}$ \\
3 & $10^{-5}\zsun$ &190 & 2370  & $3.7\times10^5$ & $4.3\times10^{17}$ & $3.9\times10^{-2}$ \\
4 & $10^{-4}\zsun$ &154 & 1730  & $3.4\times10^5$ & $3.4\times10^{18}$ & $8.6\times10^{-3}$ \\
5 & $10^{-3}\zsun$ &34 & 180 &  $1.6\times10^5$ & $8.5\times10^{19}$   & $4.3\times10^{-3}$\\
6 & $10^{-2}\zsun$ &18  & 53  & $1.0\times10^5$ & $9.1\times10^{19}$   & $2.2\times10^{-3}$\\
7 & $10^{-1}\zsun$ &20  & 62  & $1.1\times10^5$ & $1.7\times10^{20}$   & $1.1\times10^{-3}$ \\
8 & $\zsun$        &11  & 9.5  & $4.8\times10^4$ & $2.2\times10^{20}$  & $1.1\times10^{-3}$\\
\hline
\end{tabular}

\end{center}
\end{table}

\clearpage
\begin{figure}
\includegraphics[width=150mm]{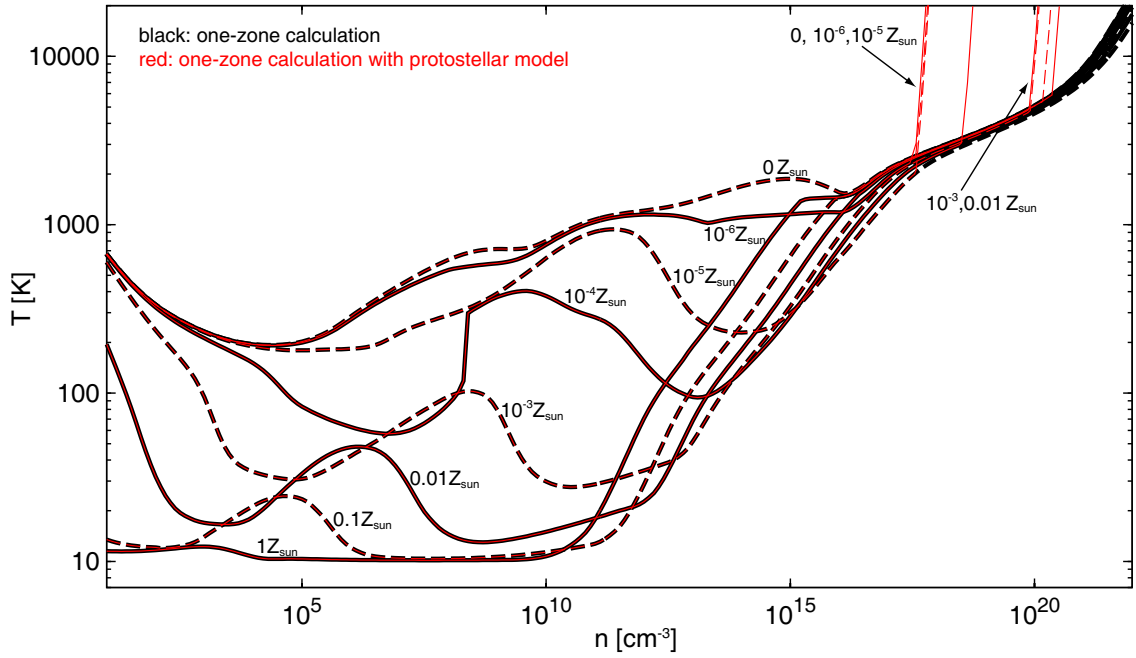}
\caption{
Temperature of collapsing gas clouds with different metallicities against the number density. 
The metallicity of each thermal evolution is indicated in the figure. 
Two evolutionary tracks are plotted for each metallicity: one given by the one-zone calculation (black) and the other by the one-zone result added  to the protostellar model in a high-density region (red).
}
\label{fig:1}
\end{figure}

\begin{figure}
\includegraphics[width=150mm]{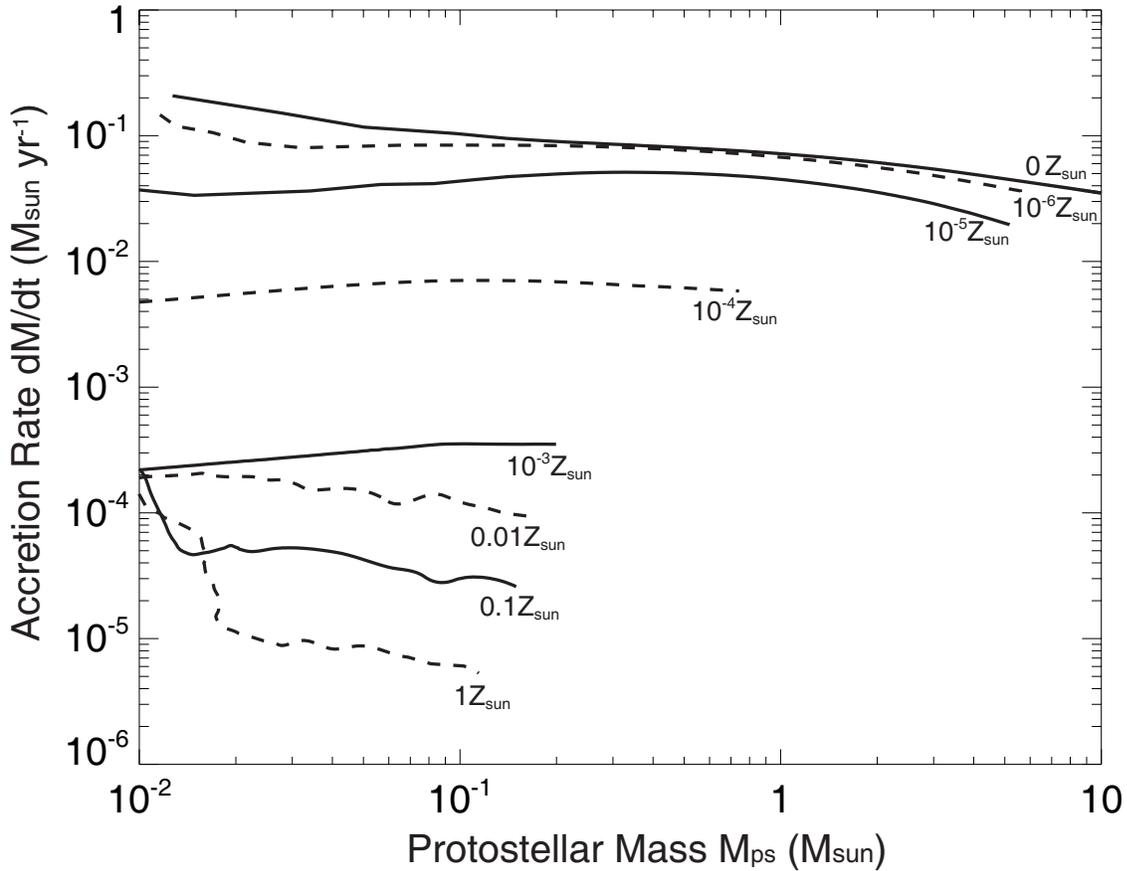}
\caption{
Mass accretion rate onto a sink versus protostellar mass.
Results are plotted for different metallicities (indicated at the right of the plots).
}
\label{fig:2}
\end{figure}

\begin{figure}
\includegraphics[width=150mm]{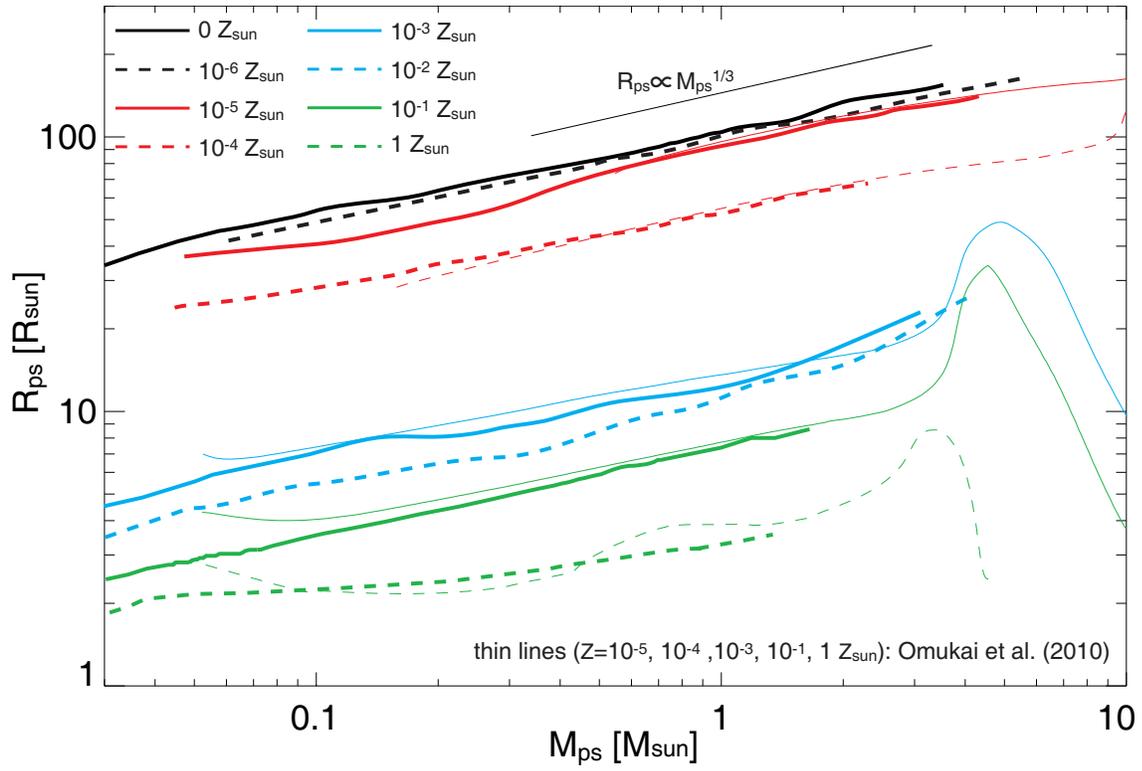}
\caption{
Protostellar radius versus protostellar mass for different metallicities.
The relation $R_{\rm ps} \propto M_{\rm ps}^{1/3}$ is also plotted. 
Protostellar radius for $Z=10^{-5}$, $10^{-4}$, $10^{-3}$, $10^{-1}$ and $1 \zsun$ calculated in \citet{omukai10} are also plotted as thin lines.
}
\label{fig:3}
\end{figure}

\begin{figure}
\includegraphics[width=150mm]{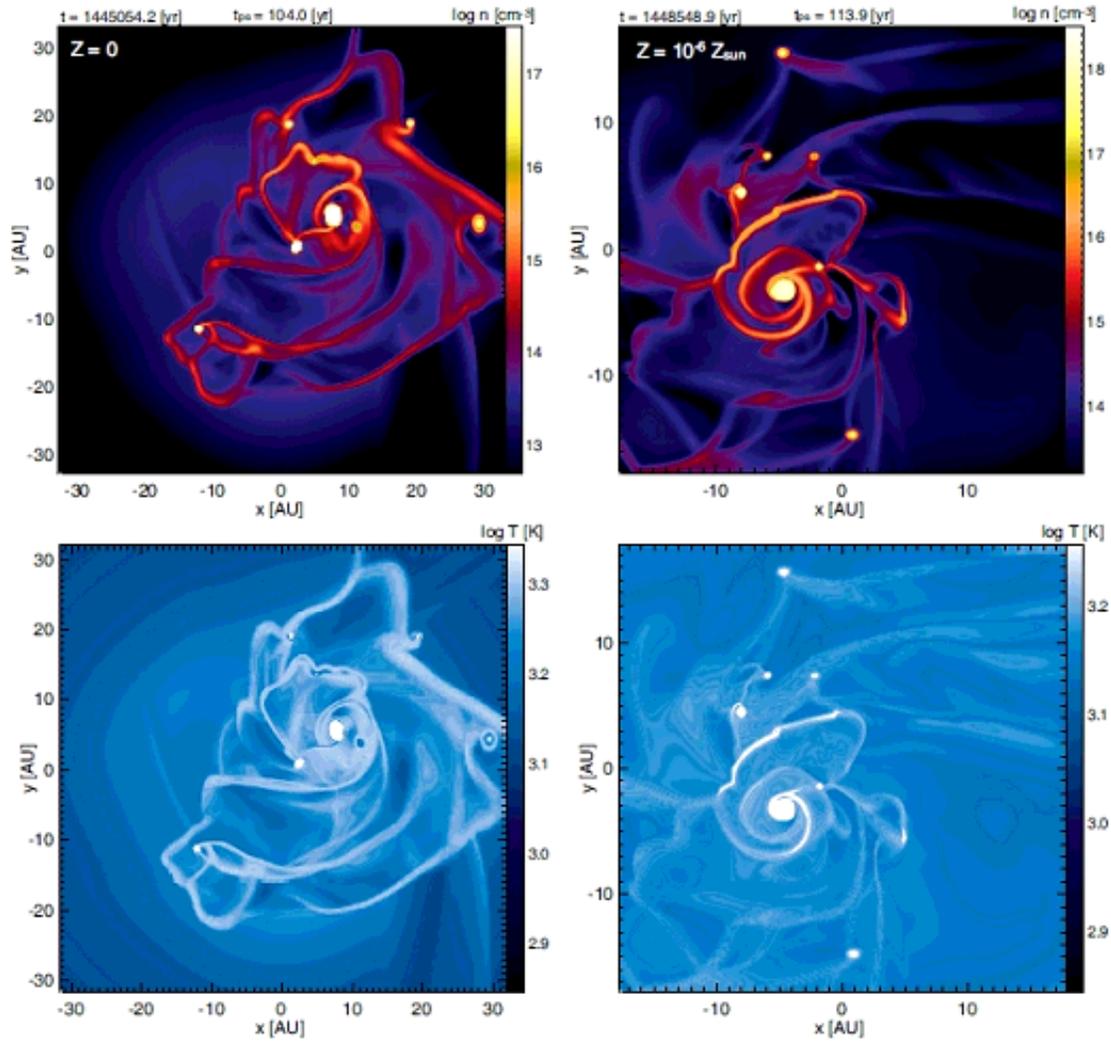}
\caption{
Density (upper panels) and temperature (lower panels) distributions on the equatorial plane in two models: $Z=0$ (left) and $10^{-6}\zsun$ (right). 
The elapsed time after the calculation starts $t$ and that after the protostar formation $t_{\rm ps}$ are described in each upper panel. 
}
\label{fig:6}
\end{figure}

\begin{figure}
\includegraphics[width=150mm]{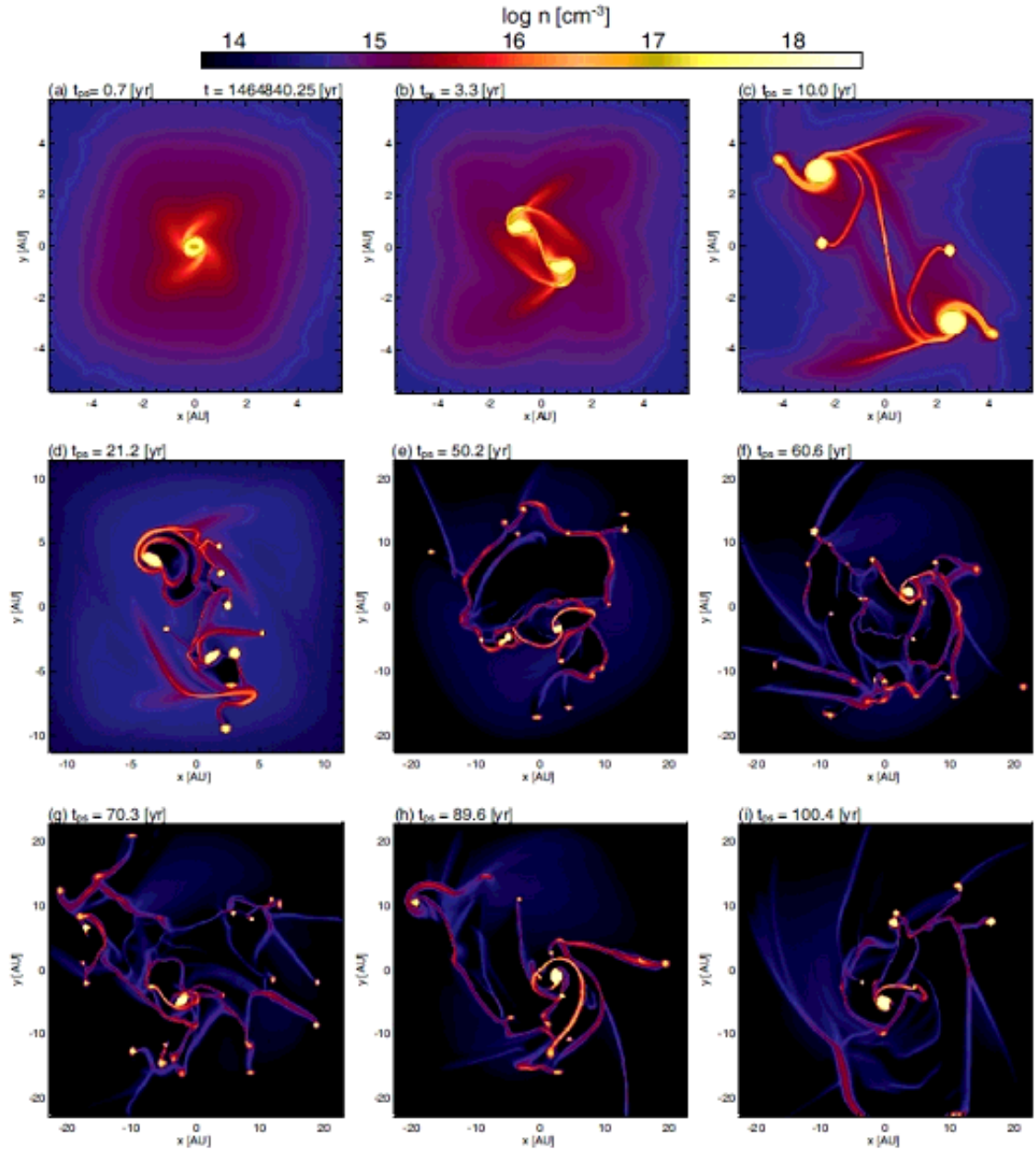}
\caption{
Time sequence of density distribution on the equatorial plane in the $Z=10^{-5}\zsun$ model.
The elapsed time after the cloud begins to collapse $t$ is described in panel ({\it a}) and that after the protostar formation $t_{\rm ps}$ is described in each panel.
The box size is different in each panel. 
}
\label{fig:7}
\end{figure}

\clearpage
\begin{figure}
\includegraphics[width=150mm]{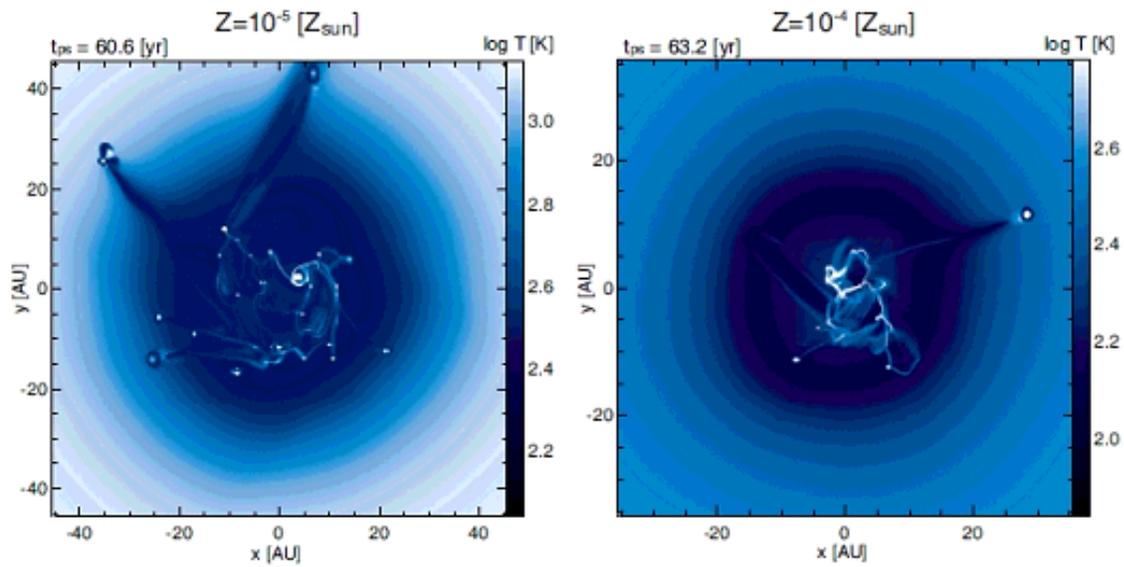}
\caption{ 
Temperature distribution in $Z=10^{-5}$ (left) and $10^{-4}\zsun$ (right). 
The elapsed time after the protostar formation $t_{\rm ps}$ is described in each panel. 
The epochs of the left and right panels correspond to Fig.~\ref{fig:7}{\it f} and Fig.~\ref{fig:8}{\it f}, respectively.
}
\label{fig:7b}
\end{figure}

\begin{figure}
\includegraphics[width=150mm]{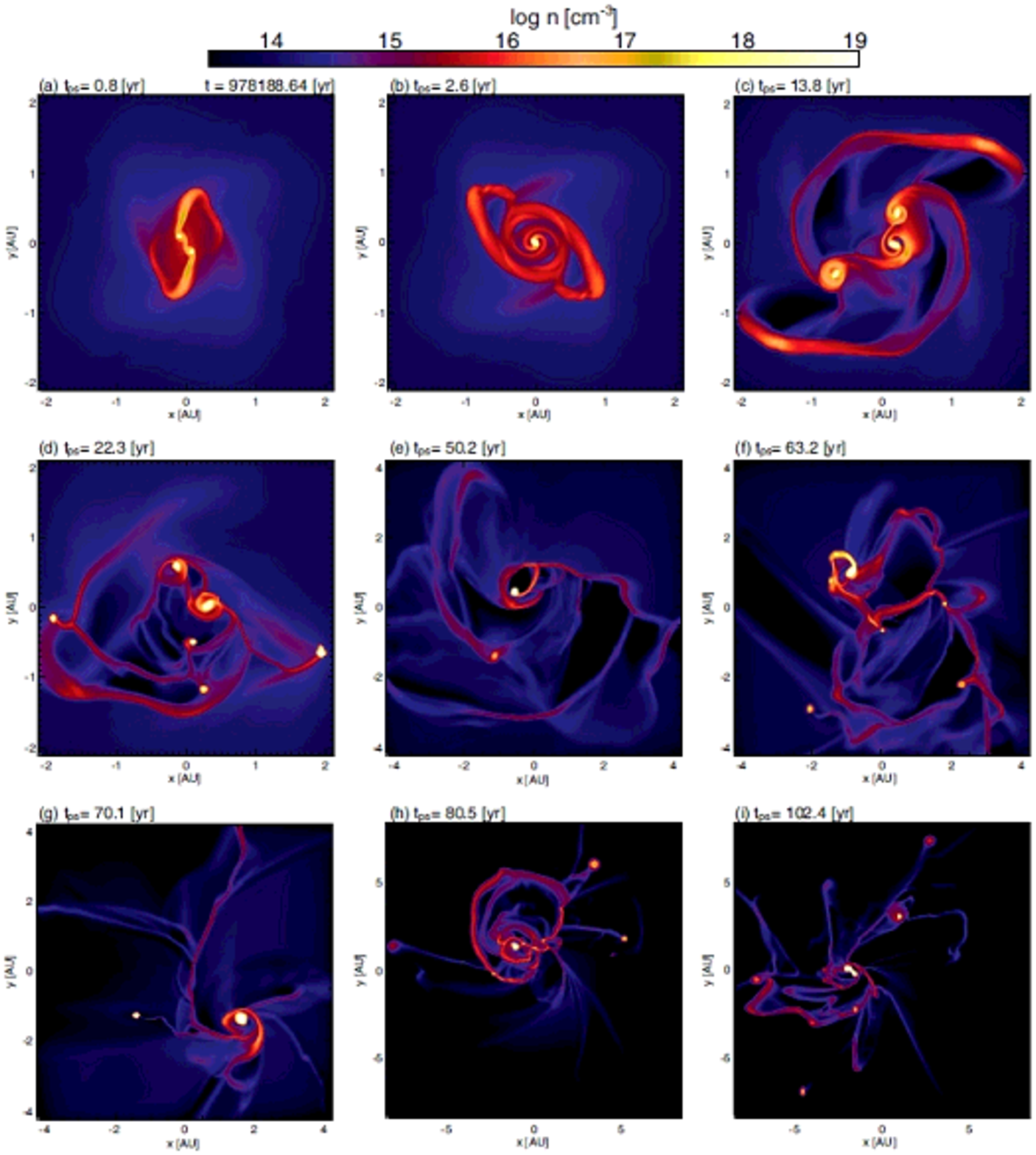}
\caption{
Same as Fig.~\ref{fig:7} for the $Z=10^{-4}\zsun$ model.
}
\label{fig:8}
\end{figure}

\begin{figure}
\includegraphics[width=150mm]{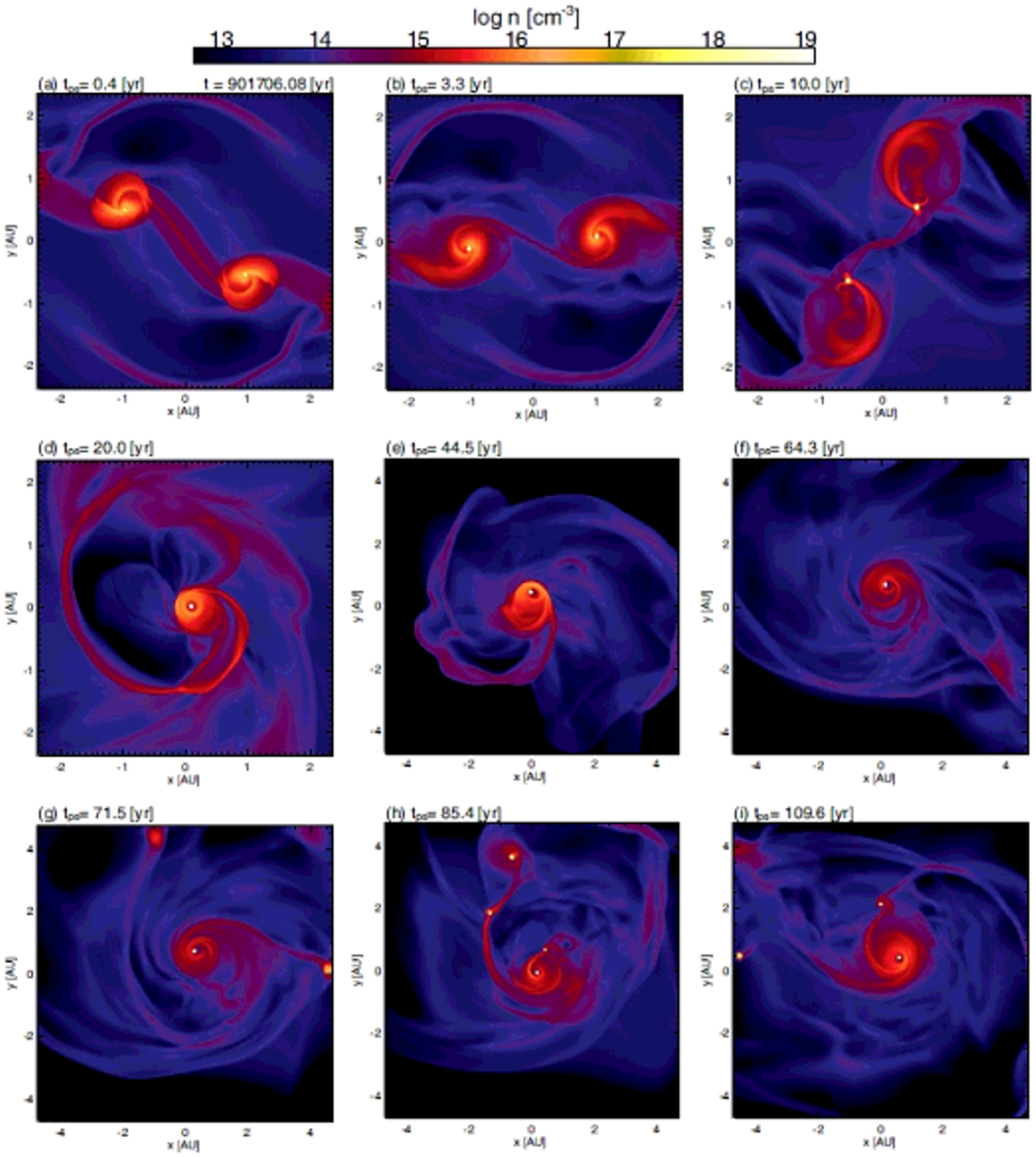}
\caption{
Same as Fig.~\ref{fig:7} for the $Z=10^{-3}\zsun$ model.
}
\label{fig:9}
\end{figure}

\begin{figure}
\includegraphics[width=150mm]{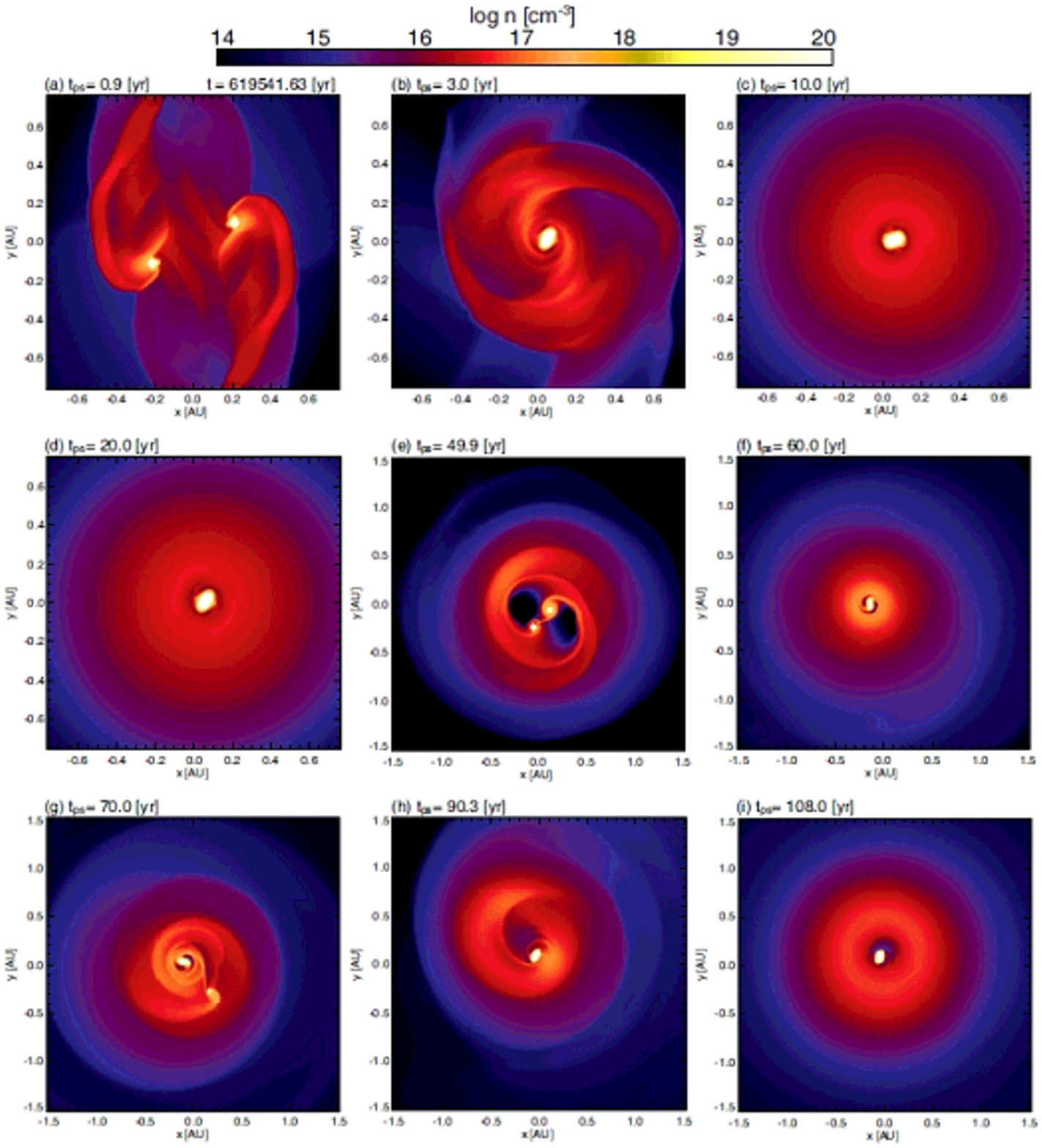}
\caption{
Same as Fig.~\ref{fig:7} for the $Z=10^{-2}\zsun$ model.
}
\label{fig:10}
\end{figure}

\begin{figure}
\includegraphics[width=150mm]{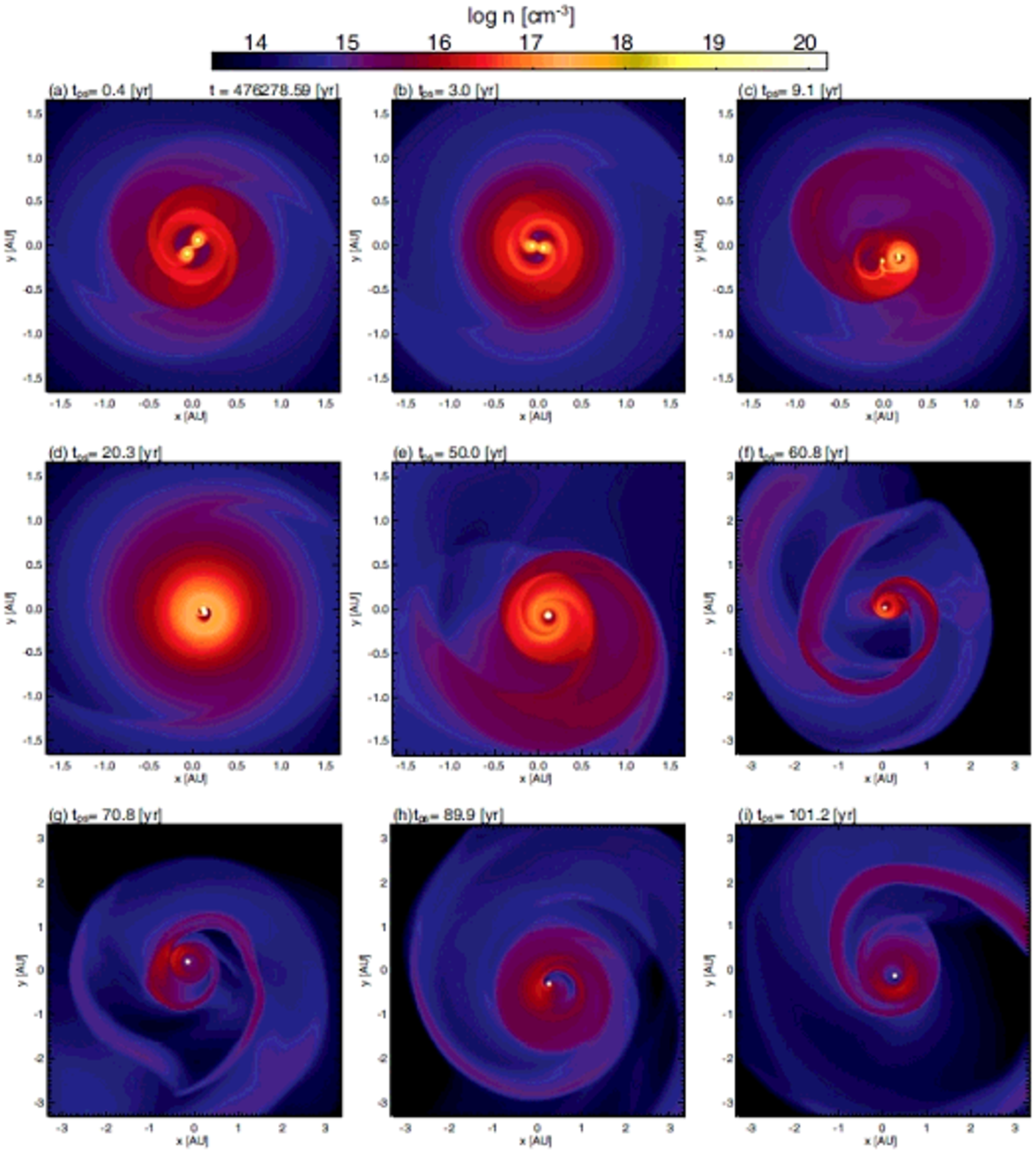}
\caption{
Same as Fig.~\ref{fig:7} for the $Z=10^{-1}\zsun$ model.
}
\label{fig:11}
\end{figure}

\begin{figure}
\includegraphics[width=150mm]{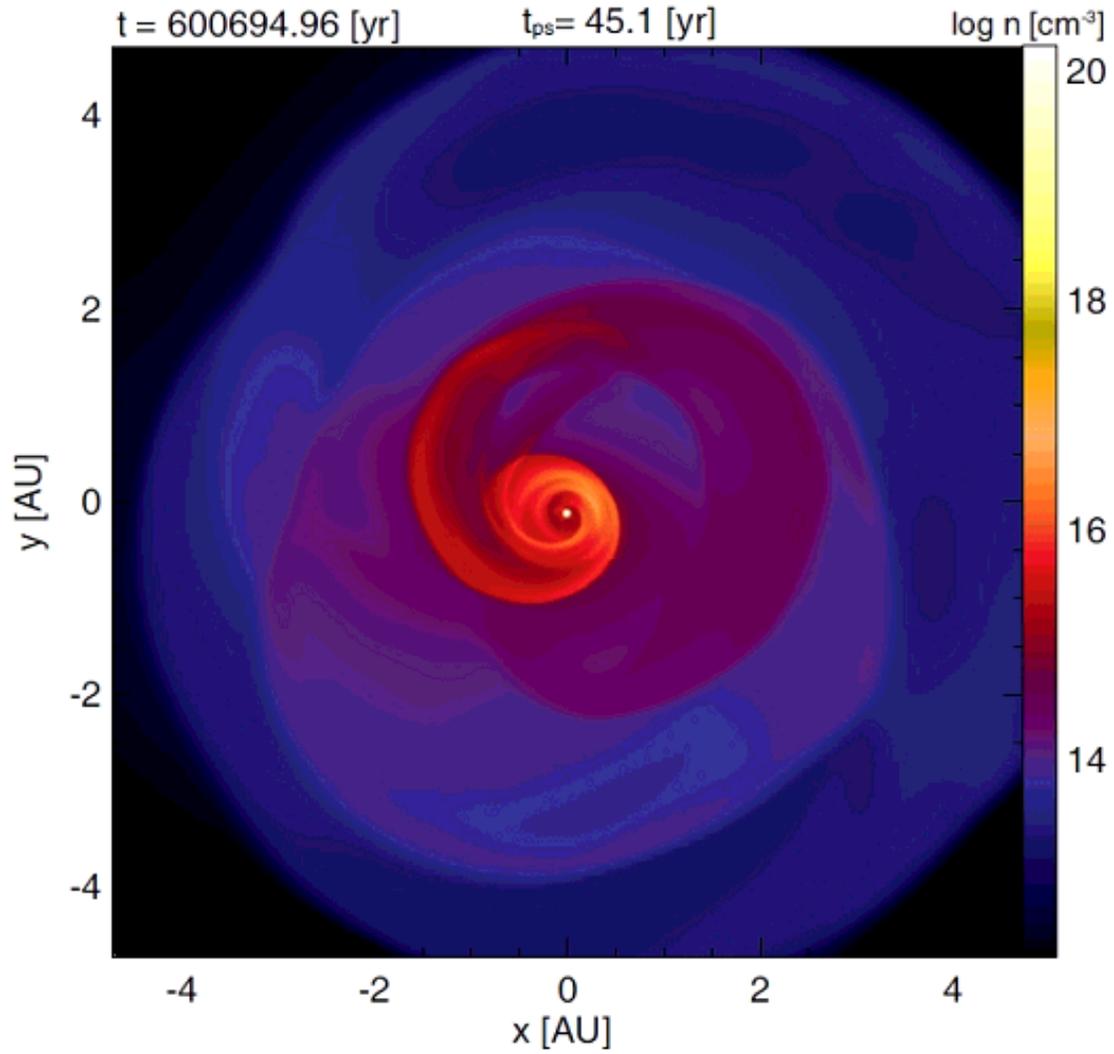}
\caption{
Density distribution on the equatorial plane in the $Z=1\zsun$ model.
The elapsed time after the calculation starts $t$ and that after the protostar formation $t_{\rm ps}$ are described.
}
\label{fig:12}
\end{figure}

\begin{figure}
\includegraphics[width=150mm]{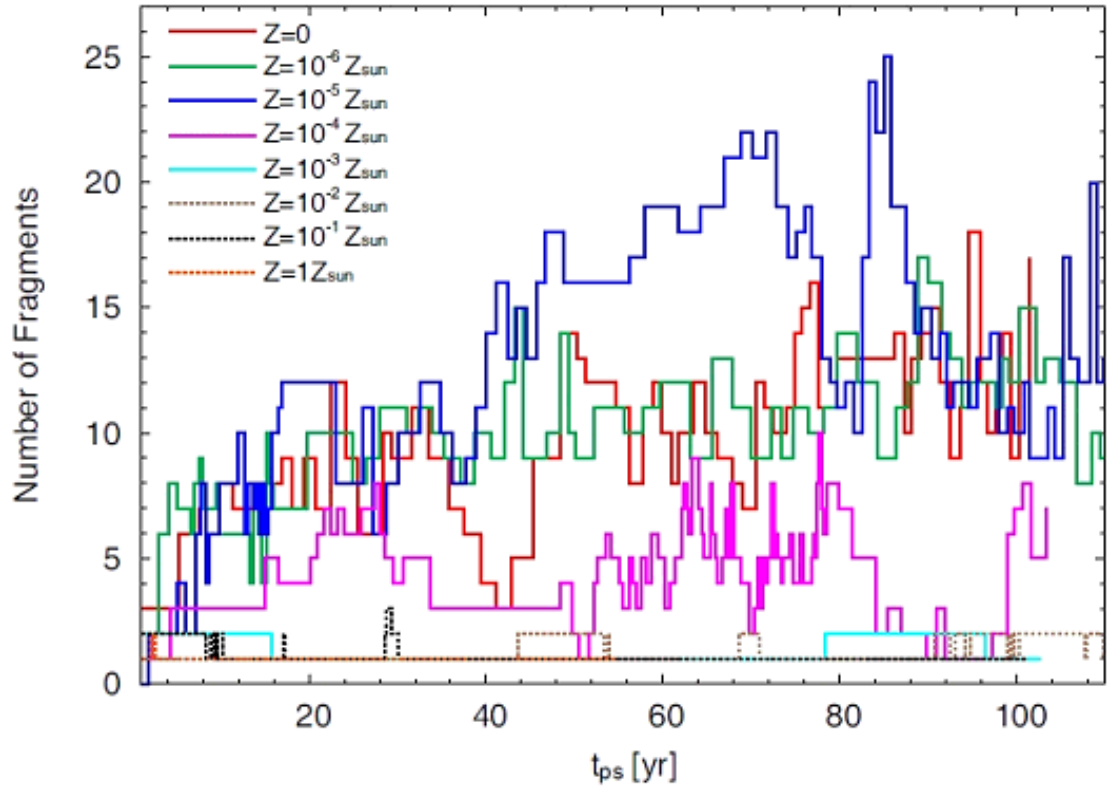}
\caption{
Number of protostars versus the elapsed time after the first protostar formation. 
Results are plotted for all models.
}
\label{fig:13}
\end{figure}

\begin{figure}
\includegraphics[width=150mm]{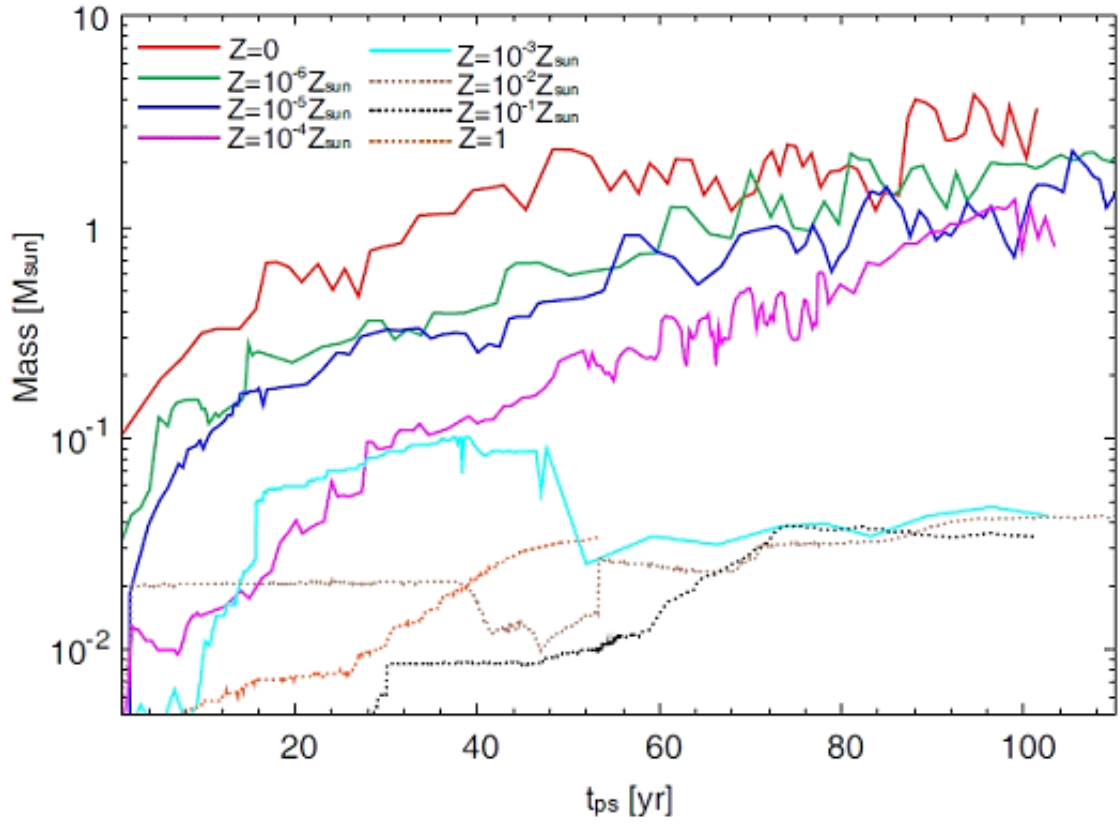}
\caption{
Mass of the most massive protostar versus the elapsed time after the first protostar formation.
Results are plotted for all models. 
}
\label{fig:14}
\end{figure}

\begin{figure}
\includegraphics[width=150mm]{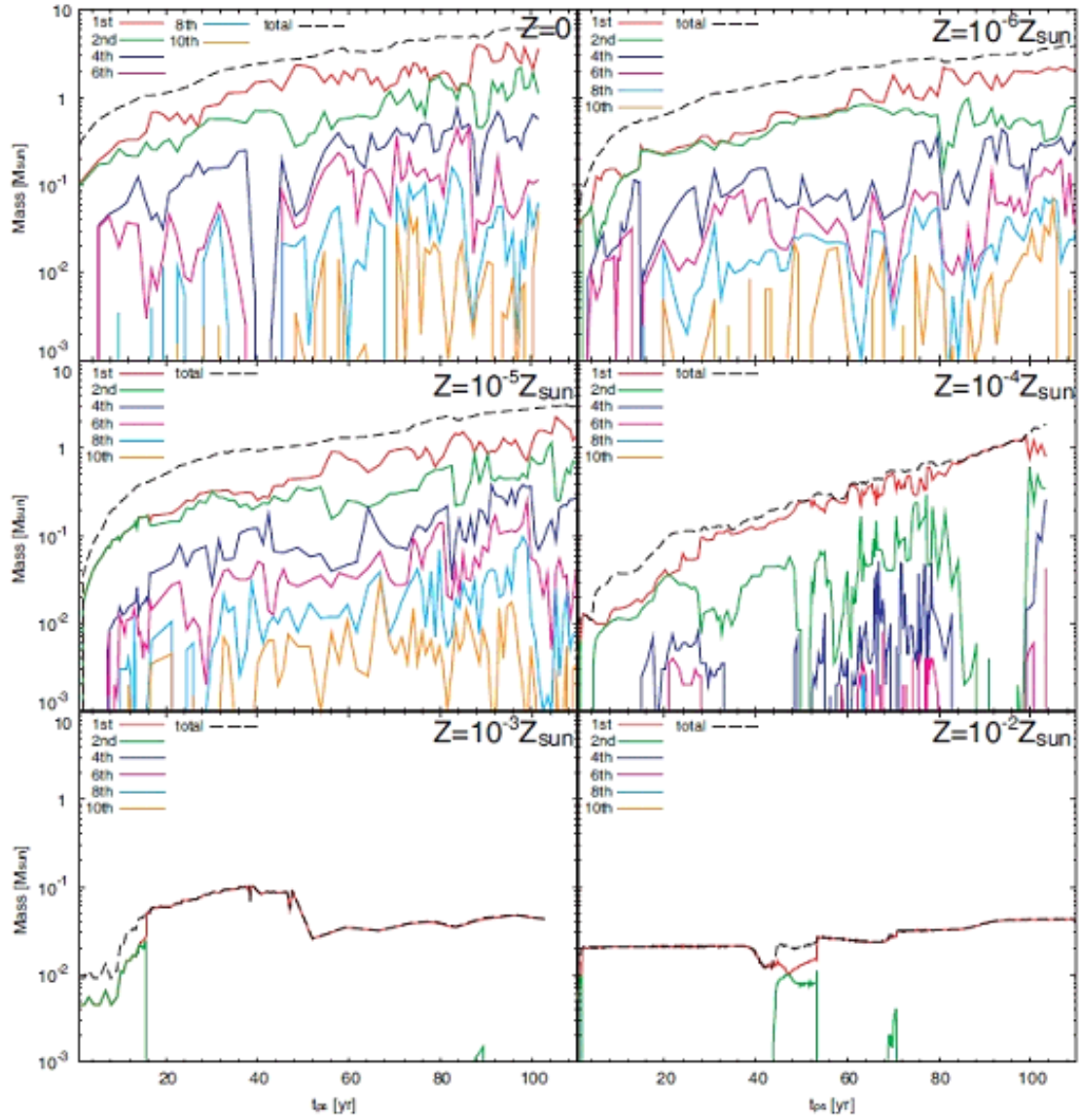}
\caption{
Mass of the first, second, fourth, sixth, eighth and tenth most massive protostars in  models $Z=0$, $10^{-6}$, $10^{-5}$, $10^{-4}$, $10^{-3}$ and $10^{-2}\zsun$ versus the elapsed time after the first protostar formation.
The total mass of protostars is also plotted by a broken line in each panel.
}
\label{fig:15}
\end{figure}

\begin{figure}
\includegraphics[width=150mm]{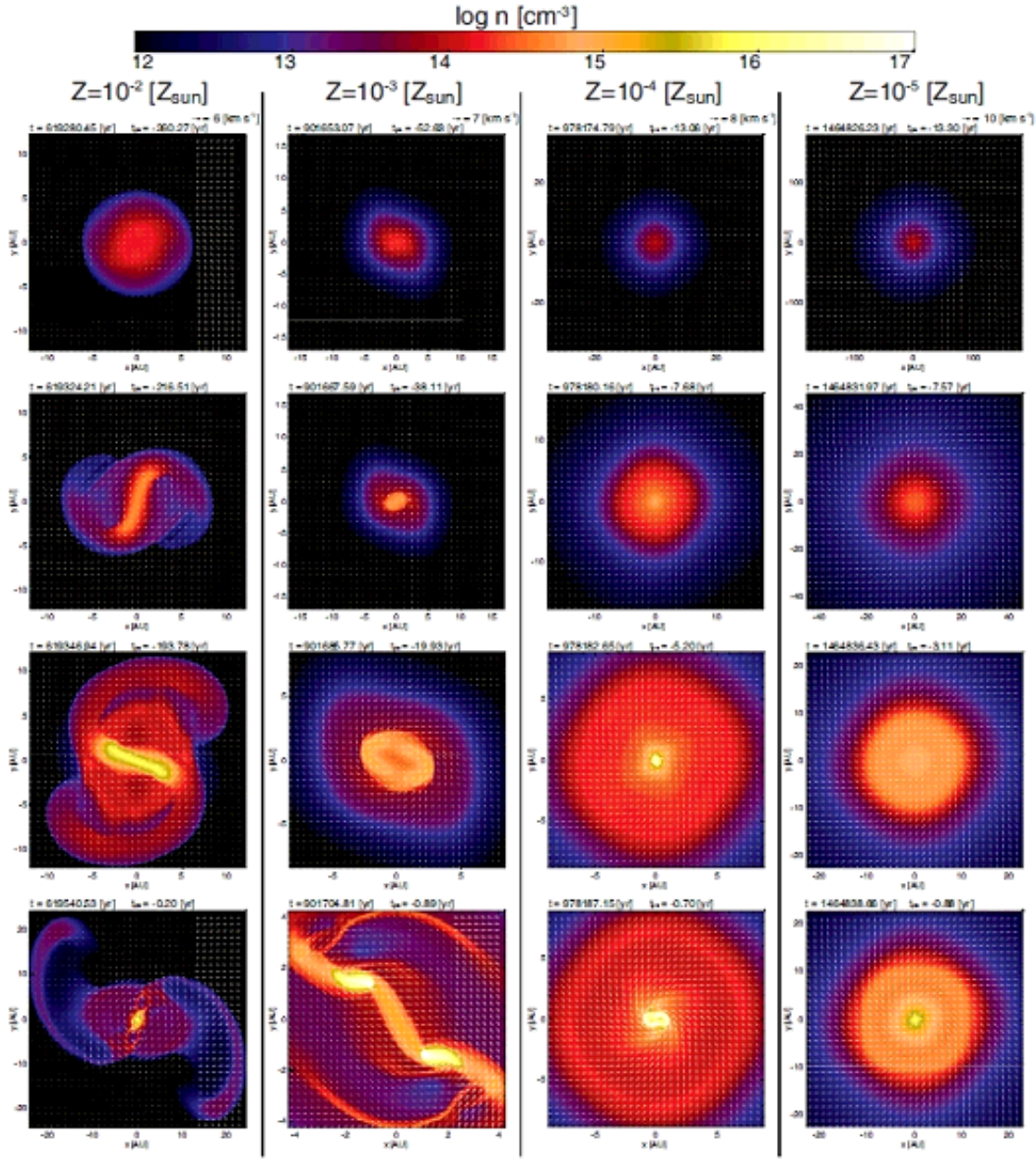}
\caption{
Time sequence (rows) of cloud evolution on the equatorial plane in the $Z=10^{-2}$, $10^{-3}$, $10^{-4}$ and $10^{-5}\zsun$ models. 
The density distribution (colour) and velocity vectors (arrows) on the equatorial plane are plotted in each panel.
The elapsed time after the cloud begins to collapse $t$ and that after the protostar formation $\tps$ are also described in each panel.
The box size is different in each panel.
}
\label{fig:16}
\end{figure}

\begin{figure}
\includegraphics[width=150mm]{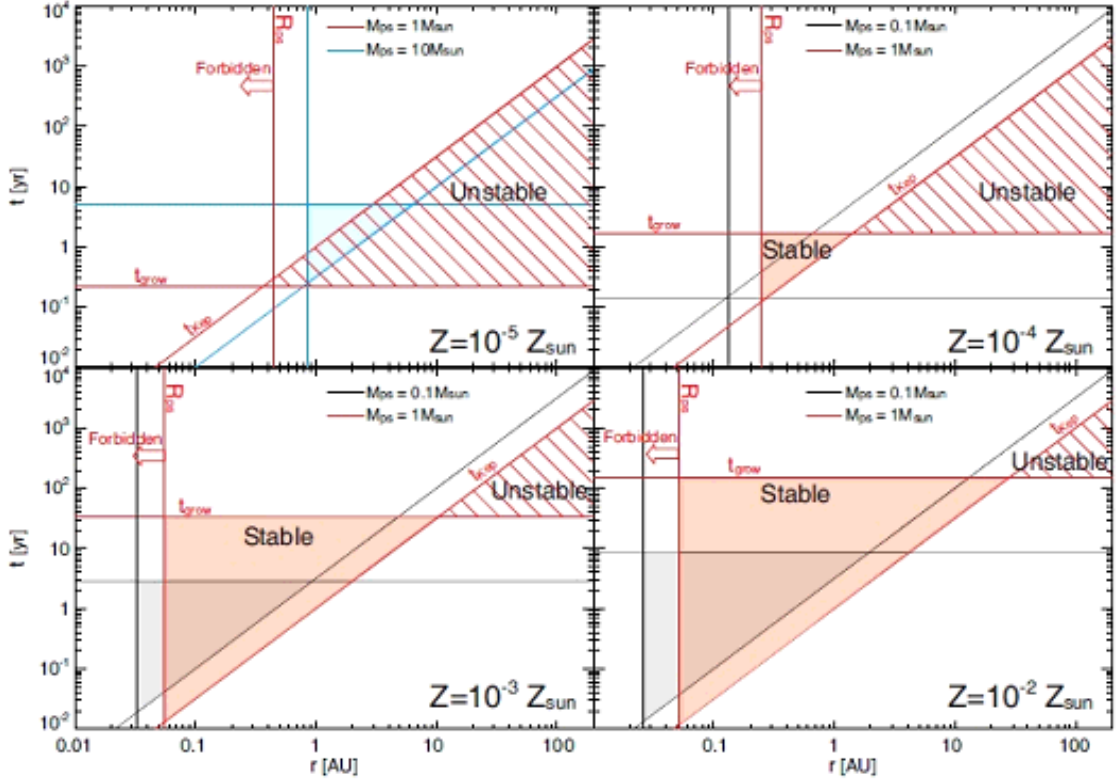}
\caption{
Disc growth $t_{\rm grow}$ and Keplerian timescale $t_{\rm Kep}$ for $M_{\rm ps}=1$ (red), $0.1$ (black) and $10\msun$ (blue) versus the radius from the protostar in the $Z=10^{-5}$, $10^{-4}$, $10^{-3}$ and $10^{-2}\zsun$ models. 
The protostellar radius $R_{\rm ps}$ in each epoch is also plotted in each panel.
The unstable and stable regions are defined as $t_{\rm grow}<t_{\rm Kep}$  and  $t_{\rm grow}>t_{\rm Kep}$ outside the protostellar radius $r>R_{\rm ps}$, respectively.
}
\label{fig:17}
\end{figure}

\begin{figure}
\includegraphics[width=150mm]{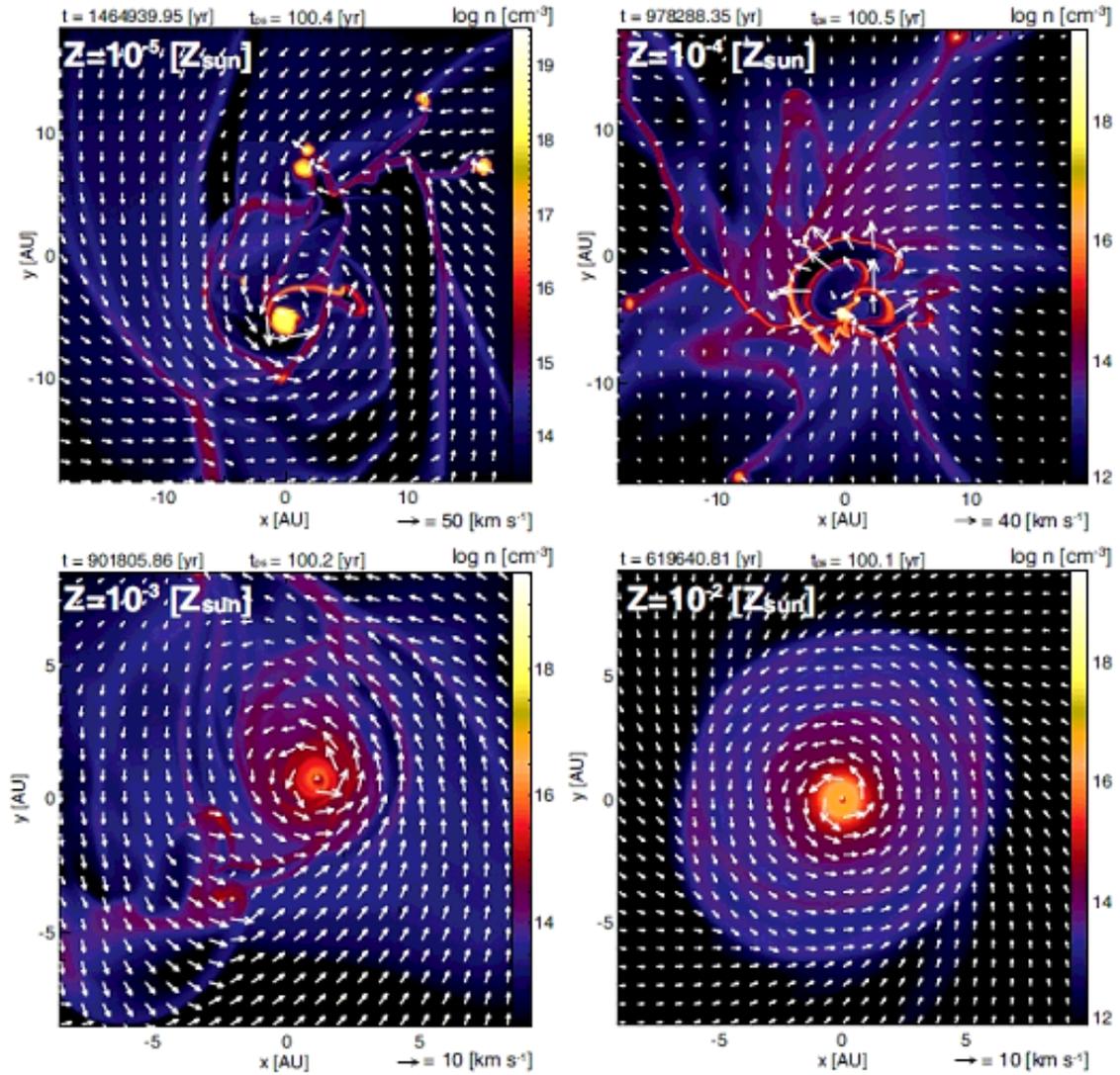}
\caption{
Density (colour) and velocity (arrow) distributions on the equatorial plane at $\tps \simeq 100$\,yr in the $Z=10^{-5}$, $10^{-4}$, $10^{-3}$ and $10^{-2}\zsun$ models. 
The elapsed time after the cloud begins to collapse $t$ and that after the first protostar formation $\tps$ are described in each panel. 
The spatial scale is different in each panel.
}
\label{fig:18}
\end{figure}

\clearpage
\appendix
\section{Test Calculation for Spherical Clouds with Protostellar Model}
\label{sec:test}
To verify our protostellar model, we calculated a non-rotating cloud with different metallicities as described in \S\ref{sec:psmodel}.
Figure~\ref{fig:4} shows the density (upper panels) and velocity (lower panels) profiles at different epochs in clouds with $Z=10^{-5}\zsun$ (left panels) and $10^{-2}\zsun$ (right panels).
The initial state for the models is a non-rotating cloud.
The protostellar surface is marked by a sudden increase in density in a high-density region. 
The figure indicates that the protostar gradually enlarges over time. 
In addition, the profiles correspond well with the analytical solutions. 
In the gas-collapsing phase prior to the protostar formation, the density is proportional to $\rho \propto r^{-2}$ in the outer region, and the radial velocity ($v_r$) gradually decreases with decreasing radius near the cloud centre \citep{larson69,omukai98}. 
On the other hand, during the gas accretion phase after protostar formation, the velocity continuously decreases with decreasing radius and suddenly becomes $v_r\simeq0$ at the protostellar surface.
The density profile has $\rho \propto r^{-1.5}$ near the protostar and $\rho \propto r^{-2}$  far from the protostar.
\citet{whitworth85} showed that  the density profile around the protostar changes from $\rho \propto r^{-2}$  to $\rho \propto r^{-1.5}$ after protostar formation \citep{larson69, shu77,hunter77}.
These features are clearly seen in  Figure~\ref{fig:4} (left panels).

Immediately before and after the protostar formation, the outer envelope is more complicated in the $Z=10^{-2}\zsun$ model than in the $Z=10^{-5}\zsun$ model.  
The red lines in the right panels of Figure~\ref{fig:4} show that, in the $Z=10^{-2}\zsun$ model, there are two shock fronts at $r\sim5\rsun$ ($\sim 0.02$\,AU) and $\sim 60\rsun$ ($\sim0.3$\,AU).
The inner shock corresponds to the protostar, whereas the outer shock corresponds to the first core or its remnant. 
In a collapsing cloud containing abundant metals and dust, after the dust cooling becomes optically thick and adiabatic heating dominates the dust cooling, the gas behaves adiabatically in the range of $10^{10}\cm \lesssim n \lesssim 10^{16}\cm$ (Fig.~\ref{fig:1}). 
During the adiabatic contraction phase, the first adiabatic core (or the first core) forms \citep{larson69,masunaga00}.
However, as seen in the right panels of Figure~\ref{fig:4}, the first core disappears several years after the protostar formation, when the star-forming cloud has no angular momentum. 
Note that when the star-forming cloud has angular momentum and contains a greater abundance of metals and dust, the first core persists and evolves into a rotation-supported disc or a circumstellar disc \citep{bate98,machida10a,bate11,bate14,machida11d,tomida13,tsukamoto13}. 
Thus, the first core plays a critical role in the star formation process.
In the $Z=10^{-5}\zsun$ model, the first core exists for a very short period  (see Fig.~3 in \citealt{machida09b}) and  immediately disappears after the protostar formation. 
The effect of the first core and cloud metallicity on the star formation is discussed in \S\ref{sec:first-core}.

Figure~\ref{fig:5} shows the time sequence of density and velocity distributions on the equatorial plane after the protostar formation in the $Z=10^{-5}\zsun$ (upper panels) and  $10^{-2}\zsun$ (lower panels) models. 
The central white region bounded by the shock front indicates the protostar. 
The figure indicates that the protostar gradually enlarges with increasing protostellar mass, as seen in Figure~\ref{fig:4}. 
In addition, the surrounding gas density gradually decreases, whereas the (negative) radial velocity increases with time. 
Figures~\ref{fig:4} and \ref{fig:5} (and Figs.~4 and 5 in \citealt{machida13a}) indicate that in the absence of angular momentum, the protostellar environment in the proximity of the protostar shows a similar evolution among models with different metallicities although the protostellar sizes (and mass accretion rates) considerably differ.
On the other hand, even in non-rotating clouds  there is a great difference in the large scale structure ($\gtrsim 0.1-1$\,AU), because the first core which is formed in a cloud with higher metallicity with a size of $\sim0.1-1$\,AU surrounds the protostar and greatly changes the circumstellar environment in the region of $\gtrsim 0.1-1$\,AU (for details, see \S\ref{sec:first-core}). 
When the cloud has angular momentum, the star formation process is largely governed by the presence of the first core and the mass accretion rate.


In this appendix, we have shown that our model approximately reproduces protostellar evolution.
In \S\ref{sec:results}, we examine the evolution of initially rotating clouds having different metallicities with the protostellar model.
Note that our modelling ignores the protostellar feedback effects of the ambient and infalling gas.
However, \citet{omukai10} pointed out that when the protostar is not very massive ($\lesssim 10-100\msun$), stellar feedback to the accretion flow is not significant yet.
This is because the high-density gas region, which is the focus of this study, has a relatively high temperature in the early star formation stage (see Fig.~21 in \citealt{omukai10}). 
Also, the effect of the radiation pressure can be ignored in the very early phase of the star formation because the protostellar luminosity is sufficiently small \citep{hosokawa09b}.
Note also that since the mass accretion rate is expected to be different in each protostar after fragmentation, the mass-radius relation shown in Figure~\ref{fig:3} may not be correctly reproduced in each fragment.   
In this study, we calculated the cloud evolution for only $\sim100$\,yr after the protostar formation, where the protostars have a mass of $< 10\msun$.
Although more detailed modelling may be necessary to correctly investigate the protostar formation and its effects, we believe that our model can qualitatively investigate the circumstellar disc formation and fragmentation process during  early star formation.

\begin{figure}
\includegraphics[width=150mm]{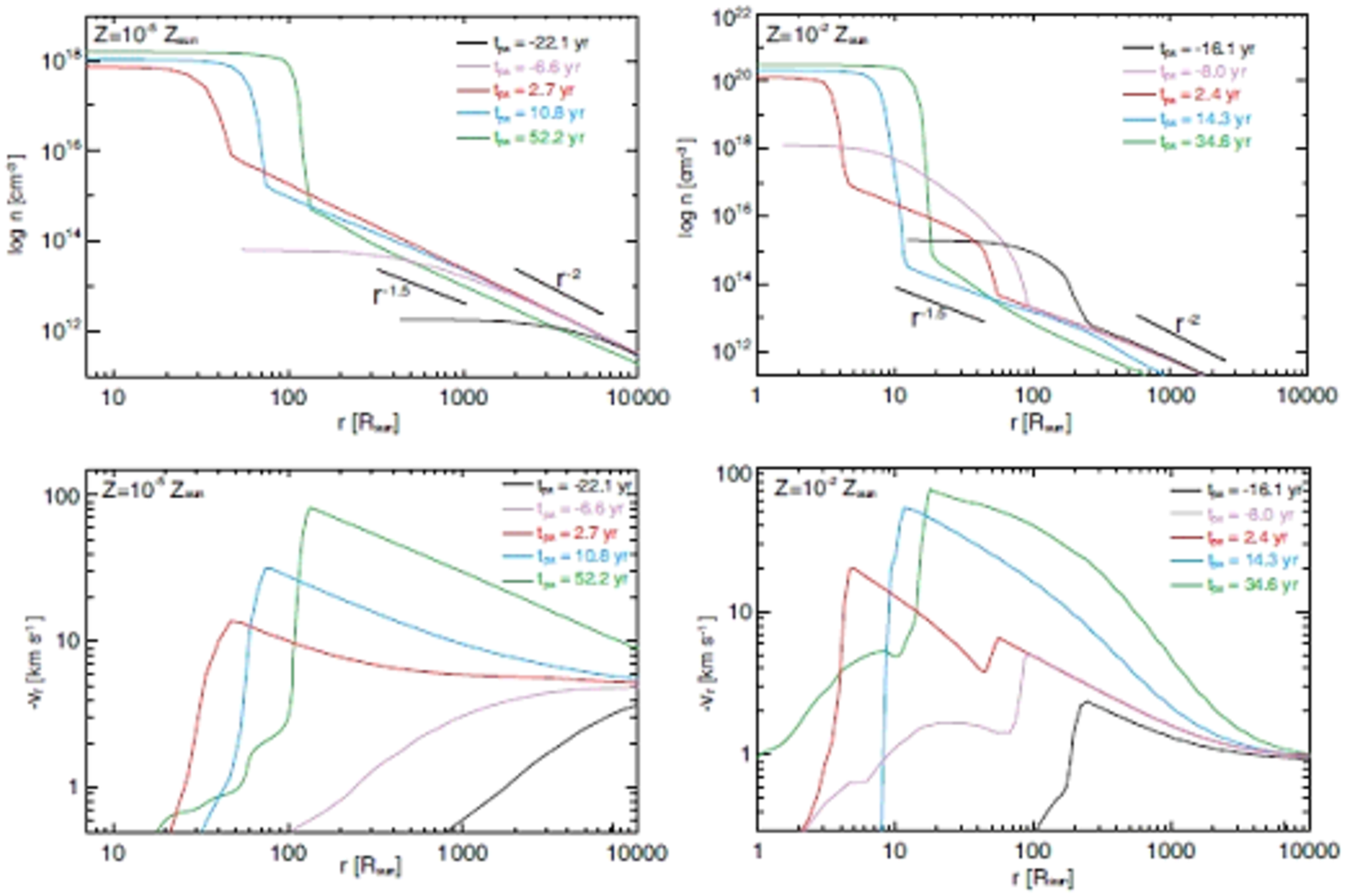}
\caption{
Density (upper panels) and velocity (lower panels) profiles at five different epochs for the models $Z=10^{-5}\zsun$ (left panels) and $10^{-2}\zsun$ (right panels) versus the radius from the cloud centre.
}
\label{fig:4}
\end{figure}

\begin{figure}
\includegraphics[width=150mm]{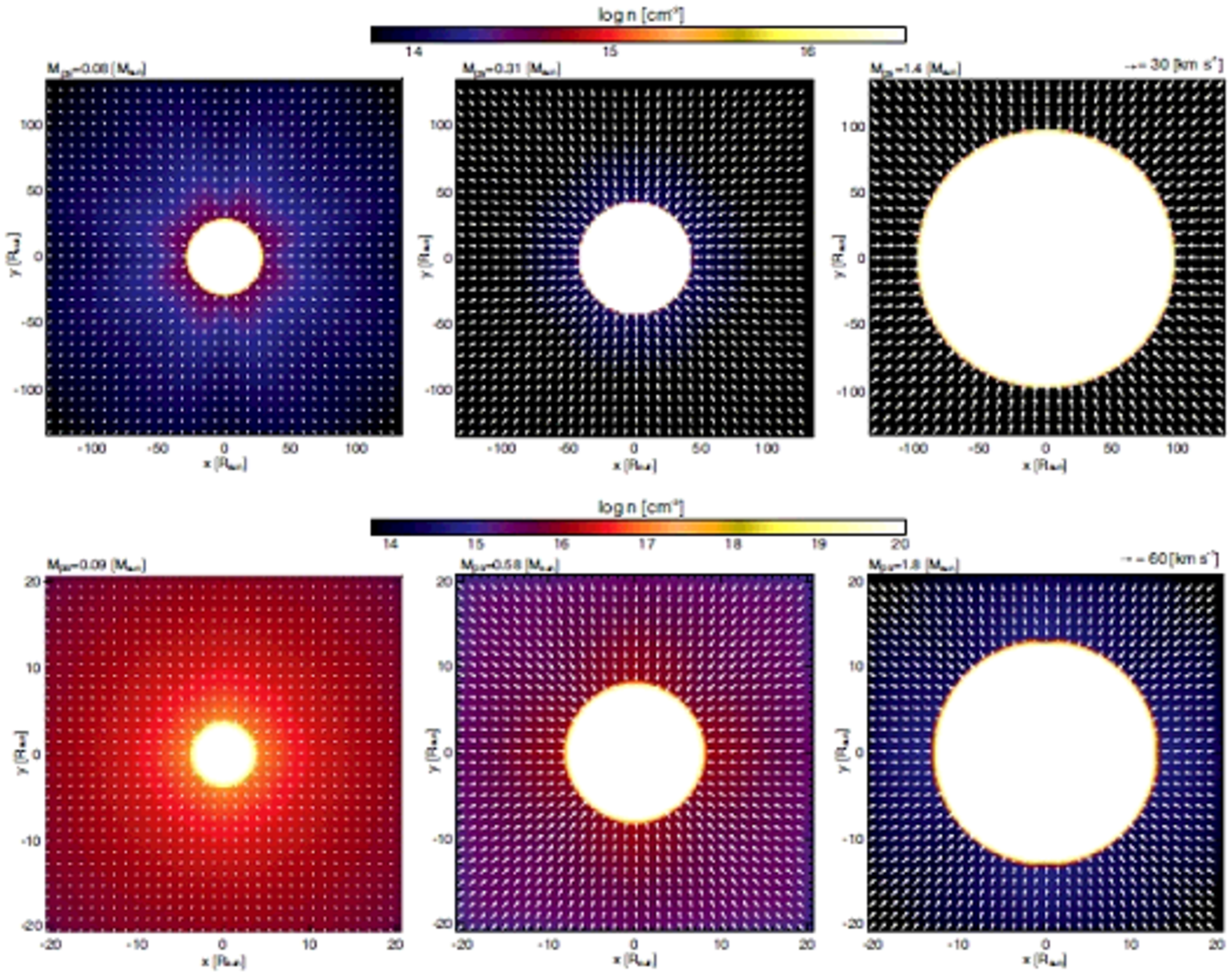}
\caption{
Time sequence of density and velocity distributions on the equatorial plane in two models: $Z=10^{-5}$ (upper panels) and $10^{-2}\zsun$ (lower panels). 
The spatial scales differ between the upper and lower panels.
The protostellar mass is described in each panel.
}
\label{fig:5}
\end{figure}

\end{document}